\pdfoutput=1
\documentclass[11pt,a4paper]{article}
\usepackage[margin=2.3cm]{geometry}
\usepackage{graphicx}
\usepackage{amsmath,amssymb}
\usepackage{booktabs}
\usepackage{tikz}
\usetikzlibrary{arrows.meta,positioning,calc}
\usepackage[font=small,labelfont=bf]{caption}
\usepackage[colorlinks=true,allcolors=blue]{hyperref}
\let\oldtabular\tabular
\renewcommand{\tabular}{\footnotesize\oldtabular}
\newif\ifdraft \draftfalse

\title{Neural variational Skyrmions as a topological benchmark: \\
emergent symmetry, precision energies and KAN--MLP comparison}
\author{Koji Hashimoto$^{1}$, Masaki Murata$^{2}$ and Gakuto Ogiwara$^{2}$\\[1.2ex]
{\small $^1$Department of Physics, Kyoto University, Kyoto 606-8502, Japan}\\
{\small $^2$Department of Information Systems, Saitama Institute of Technology, Saitama 369-0293, Japan}\\[0.6ex]
{\small E-mail: koji@scphys.kyoto-u.ac.jp, m.murata@sit.ac.jp, i5001hrr@sit.ac.jp}}
\date{\small Preprint, November 2026. KUNS-3140.}

\begin{document}
\maketitle

\begin{abstract}
We propose the three-dimensional Skyrme model as a topological benchmark for neural variational methods and construct fully three-dimensional Skyrmions with baryon number \(B=1,\ldots,4\) without imposing a symmetry ansatz, with high precision. 
For \(B=2,3,4\), the relaxation of these Skyrmions starts from \(B\) unit-charge lumps with random positions and orientations, which carry no symmetry information. The known toroidal, tetrahedral, and cubic Skyrmions nevertheless emerge spontaneously. 
The resulting variational energies for \(B=1,\ldots,4\) lie slightly below the published reference values. 
The neural method highlights the high accuracy for retaining the topological charge, 
differing from \(B\) by at most \(10^{-6}\) in the highest-precision solutions.
Our numerical method goes through physical constraint tests such as the Derrick virial relation and the Faddeev--Bogomolny bound, together with evaluation / verification through independent quadratures and  resolution sweeps.
We then compare Kolmogorov--Arnold networks (KANs) with multilayer perceptrons (MLPs) in more than a thousand training runs, both with and without a rational-map ansatz. With default settings, KANs appear markedly superior, but most of the gap disappears once the MLP activation function and learning rate are tuned. KANs remain more parameter-efficient and more robust at fixed epoch budget, while MLPs with the tuned activation function perform at least as well at fixed wall-clock time.
We also
argue that continuous gradient descent cannot smoothly change the topological sector of our neural field, clarifying the origin of topological freezing observed in our neural variational solvers.
Our results on the neural methods applied to Skyrmions pave a research ground for AI-agentic benchmarks for classical topological solitons, toward AI-assisted soliton physics.

\end{abstract}

\ifdraft\tableofcontents\newpage\fi

\section{Introduction}
\label{sec:intro}

Topological solitons are localised field configurations labelled by a
topological charge that no continuous deformation of the field can change.
They appear across physics, from nuclear
structure to cosmology and condensed matter. The prototypical
three-dimensional example is the Skyrme
model~\cite{skyrme1961,skyrme1962,manton2004}. Its field $U(\mathbf{x})$
takes values in SU(2) and carries a topological charge $B$, which is identified
with baryon number. The $B=1$ soliton models the nucleon~\cite{adkins1983}.
For $B=2,3,4$ large-scale numerical computations have shown that the
minimal-energy configurations have toroidal, tetrahedral and cubic
symmetry~\cite{manton2004,braaten1990,battye1997,battye2002}, and
solutions have since been found up to $B\sim10^2$ by relaxing carefully
designed initial configurations~\cite{feist2013}. Computing these
minimisers is hard, so an ansatz, a trial form of the field such as the
hedgehog or the rational-map approximation~\cite{houghton1998}, is often
used. The hedgehog assumes spherical symmetry, which the minimisers at
$B=2,3,4$ do not have. The rational-map ansatz is an approximation: for
$B=2$ it overshoots the energy by
$2.5\%$ (tables~9.1 and~9.2 of ref.~\cite{manton2004}).

Physics-informed neural networks (PINNs)~\cite{raissi2019,karniadakis2021}
offer an alternative. In the deep Ritz method~\cite{weinan2018} the field
itself is represented by a network and the energy functional is minimised
directly, with no restriction on the functional form. Such neural solvers have their own pitfalls, well documented for
standard PDE problems. Three matter here. First, training can suffer from
optimisation pathologies: for some problems the loss of such a solver is hard to minimise~\cite{krishnapriyan2021,wang2021}. Second, networks
that map coordinates directly to field values show a spectral bias, fitting
low frequencies first~\cite{rahaman2019,tancik2020,sitzmann2020}. The
bias matters for solitons, whose energy is concentrated in a small region. Third,
training can lower the measured energy without lowering the true one, by
making the energy density small at the points of a fixed quadrature grid and
large between them~\cite{rivera2022}.

Applied to solitons, neural solvers have been used in
symmetry-reduced settings, including electroweak Skyrmions through
one-dimensional profiles~\cite{criado2021}, brane
Skyrmions~\cite{brane2026} and families of boson stars~\cite{liu2026boson}.
Neural solvers have also been used for minimal surfaces in
curved spacetime, where the network takes one or two
coordinates~\cite{hashimoto2026minimal}.
To our knowledge, however, no neural network has been used to solve the
standard Skyrme model in three dimensions without a symmetry ansatz.
In addition, Kolmogorov--Arnold networks (KANs)~\cite{liu2024kan} have not
been applied to topological solitons in three or more spatial dimensions.
In particular, KANs are especially intriguing from a field-theory perspective because their learnable one-dimensional functions may expose hidden analytic structure in Skyrmion configurations, opening a possible route toward exact or quasi-exact soliton solutions.

At the same time, the question of which network to use is unsettled.
KANs~\cite{liu2024kan,liu2024kan2} are motivated by the Kolmogorov--Arnold
superposition theorem~\cite{kolmogorov1957}. They replace fixed activation
functions by learnable splines on the edges of the network. A rapidly
growing literature (see~\cite{somvanshi2024} for a survey) has adapted them
to physics-informed training~\cite{shukla2024,kinn2024,rigas2024}, domain
decomposition~\cite{howard2024} and neural differential
equations~\cite{koenig2024}. On PDE benchmarks KANs outperform multilayer
perceptrons (MLPs) in some studies and underperform in
others~\cite{shukla2024,kinn2024,unified2026}.
The literature is contradictory.

The Skyrme model provides a benchmark of a different character:
\begin{enumerate}
\item the field is \emph{manifold-valued} ($S^3$), enforced by construction;
\item solutions carry a discrete \emph{topological charge} that a correct
      method must quantise and preserve;
\item the Faddeev--Bogomolny bound $E\ge|B|$ (in units of $12\pi^2$) and the
      Derrick virial identity $E_2=E_4$ give exact validity checks;
\item high-precision energies from three decades of classical
      numerical work are available for comparison. Battye and
      Sutcliffe~\cite{battye2002,battye2001} computed the energies at
      $B=1$--$4$ that are listed in table~\ref{tab:bounds}. We call these
      four energies the reference energies.
\end{enumerate}
We should be clear about the reasons for using the neural approach. For the Skyrme
model a grid-based minimisation reaches a solution in
minutes~\cite{gudnason2022}, so speed is not a reason to use a network. There are two reasons. First, a network needs no mesh, no ansatz, no assumed symmetry and no hand-made initial data to represent the field.
The last three are choices that could change the result of a comparison between networks.
Leaving out these three choices makes the model usable as an architecture benchmark. Second, the method transfers to other problems. The same method applies
to generalised Skyrme models, higher charges and other target
manifolds, because it assumes no symmetry.

We obtain full three-dimensional solutions for $B=1,2,3,4$ without imposing
any symmetry. At $B=2,3,4$ the runs start from random multi-lump initial data, and the
known symmetries emerge on their own (figure~\ref{fig:iso}). The converged
energies are variational upper bounds, and the upper bounds lie slightly
below the published reference values~\cite{battye2002,battye2001}. At $B=1$ the spherically symmetric
reduction can be solved to high precision. Our $B=1$ energy agrees with the
solution of the reduction, while the published value lies $0.06\%$ above the
same solution. On this benchmark we
compare KANs and MLPs in more than a thousand runs, both with and without the rational-map ansatz, closing the fairness gaps one by one. With default settings KANs appear far superior, but once
the activation function and learning rate are tuned the two approach the same
energy at a similar cost. We argue that much of the disagreement in the KAN-versus-MLP literature comes from how networks are tuned. We also explain topological freezing: why gradient descent cannot change the topological charge of a compactified network
field except where the field becomes singular. The field should therefore already carry the target charge when
the energy minimisation starts. We show that more runs succeed when the
network fits the initial data more accurately.

The paper is organised as follows. Section~\ref{sec:model} defines the model.
Section~\ref{sec:methods} describes the field representations, the
variational loss and the randomised quadrature, the architectures, the
imprinting initialisation, and the verification protocol used to audit every run. Section~\ref{sec:fair} presents the controlled
architecture comparison, section~\ref{sec:free} the ansatz-free solutions,
their verified variational energies and the architecture comparison
repeated in that setting, and section~\ref{sec:freeze} the analysis of
topological charge under training. Section~\ref{sec:discussion} discusses implications, limitations and outlook.
Section~\ref{sec:conclusion} summarises the results.

\section{The Skyrme model as a variational benchmark}
\label{sec:model}

The static energy of the Skyrme model, in the standard energy and length
units of ref.~\cite{manton2004}, is
\begin{equation}
\begin{aligned}
E[U]&=E_2+E_4,\\
E_2&=-\frac{1}{12\pi^2}\int \tfrac12\,\mathrm{Tr}\,(R_iR_i)\,\mathrm{d}^3x,\\
E_4&=-\frac{1}{12\pi^2}\int \tfrac{1}{16}\,
\mathrm{Tr}\,\big([R_i,R_j][R_i,R_j]\big)\,\mathrm{d}^3x,
\end{aligned}
\label{eq:energy}
\end{equation}
where $R_i=(\partial_iU)U^\dagger$ is the $su(2)$-valued current of the
field $U(\mathbf{x})\in\mathrm{SU}(2)$ at $\mathbf{x}\in\mathbb{R}^3$, the
integrals run over all of $\mathbb{R}^3$, $\mathrm{Tr}$ is the trace over
the $2\times2$ matrix indices, and repeated Latin indices $i,j=1,2,3$ are
summed. The subscripts count spatial derivatives: $E_2$ is quadratic and
$E_4$ quartic. The factor $1/12\pi^2$ is the normalisation of
ref.~\cite{manton2004}, chosen so that the bound~\eqref{eq:fb} below reads
$E\ge|B|$.

Finite energy requires $U$ to tend to a constant matrix as
$|\mathbf{x}|\to\infty$. Following ref.~\cite{manton2004} we impose
$U\to 1_2$, the $2\times2$ identity, which fixes the vacuum and breaks the
chiral symmetry $U\mapsto\mathcal{O}_1U\mathcal{O}_2$, with constant
$\mathcal{O}_1,\mathcal{O}_2\in\mathrm{SU}(2)$, to the isospin symmetry
$U\mapsto\mathcal{O}U\mathcal{O}^\dagger$. This boundary condition
compactifies the domain to $S^3=\mathbb{R}^3\cup\{\infty\}$, so that $U$ is
a map $S^3\to S^3$. Since $\pi_3(S^3)=\mathbb{Z}$, such maps fall into
classes labelled by an integer, the degree
\begin{equation}
B=-\frac{1}{24\pi^2}\int\varepsilon_{ijk}\,
\mathrm{Tr}\,(R_iR_jR_k)\,\mathrm{d}^3x\;\in\;\mathbb{Z},
\label{eq:degree}
\end{equation}
Here $\varepsilon_{ijk}$ is the totally antisymmetric Levi-Civita symbol
with $\varepsilon_{123}=1$. The degree is invariant under any continuous
deformation of the field and is identified with baryon
number~\cite{manton2004}.

For computation we use the equivalent strain-tensor form of
refs.~\cite{manton2004,manton1987}. Write $U=q_0 1_2+i\,q_a\tau_a$, where
$\tau_a$ are the Pauli matrices and $a=1,2,3$; the four real components
$q=(q_0,q_1,q_2,q_3)$ satisfy $q\cdot q=1$, the dot denoting the Euclidean
inner product on $\mathbb{R}^4$, so the field is equivalently a map
$\mathbf{x}\mapsto q\in S^3\subset\mathbb{R}^4$, with the boundary condition
$U\to 1_2$ becoming $q\to q_{\rm vac}=(1,0,0,0)$ at spatial infinity. The strain tensor
$D_{ij}=-\tfrac12\mathrm{Tr}(R_iR_j)=\partial_iq\cdot\partial_jq$ is a
symmetric $3\times3$ matrix, and the energy density $e$ and the baryon
density $b$ are
\begin{equation}
e=\frac{1}{12\pi^2}\Big[\mathrm{tr}\,D+\tfrac12\big((\mathrm{tr}\,D)^2
-\mathrm{tr}\,D^2\big)\Big],\qquad
b=-\frac{1}{2\pi^2}\det\big[q,\partial_1q,\partial_2q,\partial_3q\big],
\label{eq:densities}
\end{equation}
with $E=\int e\,\mathrm{d}^3x$ and $B=\int b\,\mathrm{d}^3x$. Here
$\mathrm{tr}$ is the trace of the $3\times3$ matrix $D$. In $b$, the four
vectors $q,\partial_1q,\partial_2q,\partial_3q$ form the columns of a
$4\times4$ matrix, and $\det$ is its determinant. Up to sign it is the
Jacobian of the map $\mathbf{x}\mapsto q$, so $b$ measures how much of the
target $S^3$, of volume $2\pi^2$, is covered per unit volume, and $B$ counts
how many times it is covered.

Two exact relations constrain every true solution. First, the
densities~\eqref{eq:densities} satisfy $e\ge|b|$ at every point, which
follows from an inequality among the eigenvalues of $D$~\cite{manton2004}.
Integrating gives the Faddeev--Bogomolny
bound~\cite{faddeev1976,bogomolny1976}
\begin{equation}
E\;\ge\;|B|.
\label{eq:fb}
\end{equation}
The bound is not attained for any $B\neq0$; the $B=1$ Skyrmion lies $23\%$
above it~\cite{manton2004}. It is therefore a weak bound on the energy but
an absolute one: a measured value $E<|B|$ signals a numerical error, not a
good solution, and we use it as a hard rejection criterion. Second, under the
rescaling $q_\lambda(\mathbf{x})=q(\mathbf{x}/\lambda)$ the two terms scale
oppositely, $E[q_\lambda]=\lambda E_2+\lambda^{-1}E_4$. Stationarity at
$\lambda=1$ (Derrick's scaling argument~\cite{derrick1964}) gives the virial
identity
\begin{equation}
E_2=E_4
\label{eq:virial}
\end{equation}
at any static solution. The ratio $E_2/E_4$ is therefore a convergence diagnostic that requires no
extra computation.

We therefore propose the Skyrme model not merely as another nonlinear PDE test, but as a
topological benchmark for scientific learning methods. Its significance is that success is judged
by structures that are intrinsic to the field theory itself: the solution must lie in the correct
topological sector, respect exact energetic and virial constraints, and approach a nonperturbative
soliton whose energy can be compared with decades of numerical work. In this sense the benchmark
tests more than function approximation; it asks whether a computational method can preserve and
discover the nonperturbative structure of a nonlinear field theory. We expect such domain-physics
benchmarks to become increasingly important as AI systems and autonomous agents are used for
scientific computation. Progress should then be measured not only on generic numerical tasks,
but on problems where the target is a genuine question in theoretical physics and where the
answer can be audited by exact physical structure. Topological solitons provide a particularly
sharp setting for this purpose, because an apparently accurate numerical field can still be
physically wrong if it carries the wrong degree, violates a bound, or fails a virial identity.
The present work reaches only low baryon number, but the same benchmark naturally extends to
higher-charge sectors, where one can ask for increasingly precise energies, for the spontaneous
emergence and competition of nontrivial symmetries, and eventually for the semiclassical fluctuation
spectrum around the resulting Skyrmions. We view these as natural community benchmarks for
future neural and AI-assisted field-theory calculations, and hope that they will provide common
targets on which increasingly capable methods can be compared.

\section{Methods}
\label{sec:methods}

\subsection{Neural field representation}
\label{sec:field}
Networks take the compactified coordinates
$\mathbf{u}=\mathbf{x}/(1+r)$, with $r=|\mathbf{x}|$, as input. We use
spherical polar coordinates $(r,\theta,\phi)$ for $\mathbf{x}$ throughout.
The map $\mathbf{x}\mapsto\mathbf{u}$ takes
$\mathbb{R}^3$ onto the open unit ball $\mathbb{B}^3$ and spatial infinity
onto its boundary, so the whole field lives on a bounded domain. Each component of $\mathbf{u}$ lies in $(-1,1)$, inside the grid $[-1,1]$ of the KAN splines (section~\ref{sec:arch}), and $(-1,1)$ is also a common range for the inputs of an MLP. The
network outputs a raw four-vector $\mathbf{v}(\mathbf{u})\in\mathbb{R}^4$,
which the normalisation map $\mathcal{N}[\mathbf{v}]=\mathbf{v}/\|\mathbf{v}\|$
projects onto the target manifold. The constraint $q\in S^3$ therefore
holds exactly by construction, not through a penalty. We use two field
constructions, which we call the free and anchored tracks.
\paragraph{Free track.} $q=\mathcal{N}\!\big[\mathbf{v}(\mathbf{u})\big]$. The
far-field vacuum and the topological sector are not built in; they have to
come from the loss and the initialisation
(section~\ref{sec:freeze}). We use the term
ansatz-free in a precise sense: no symmetry reduction, no background field
and no functional form of the solution is imposed.
\paragraph{Anchored track.}
$q=\mathcal{N}\!\big[q_{\rm base}+g(r)\,\mathbf{v}(\mathbf{u})\big]$, where the network
corrects a fixed background, the rational-map ansatz~\cite{houghton1998}
\begin{equation}
q_{\rm base}(\mathbf{x})
=\big(\cos f(r),\;\sin f(r)\,\hat{\mathbf{n}}_R(\theta,\phi)\big).
\label{eq:rmap}
\end{equation}
Here
$\hat{\mathbf{n}}_R=\big(2\,\mathrm{Re}\,R,\;2\,\mathrm{Im}\,R,\;1-|R|^2\big)/(1+|R|^2)$
is the unit vector induced by the rational map $R(z)=z^B$ of the
stereographic angular coordinate $z=\tan(\theta/2)\,e^{i\phi}$. The anchored
track uses only $B=1$ and $2$, where $z^B$ minimises the energy within the
rational-map approximation at that degree. Section~\ref{sec:fair} compares
the architectures at $B=2$ only. The $B=1$ runs are used in the appendices:
for the charge-preservation check (appendix~\ref{app:charge}), the
implementation comparison against \texttt{pykan}
(appendix~\ref{app:implcost}), and the quadrature failure of the
\texttt{FastKAN} variant of section~\ref{sec:arch}
(appendix~\ref{app:quadfail}). The background $q_{\rm base}$
therefore
carries the topological charge $B$ and the boundary conditions exactly. The
radial profile $f(r)$ is pre-optimised by solving the one-dimensional
reduction with a small auxiliary network, imposing $f(0)=\pi$ and the
power-law tail $f\sim r^{-p}$ with $p=\tfrac12(1+\sqrt{1+8B})$, the decaying
solution of the equation for $f$ linearised at large $r$. The
envelope $g(r)=4s(1-s)^3$ with $s=r/(3+r)$ vanishes linearly at $r=0$ and as
$r^{-3}$ at infinity.\footnote{This envelop function enforces that at $r=0$ the resultant profile needs to be equal to $q_{\rm base}(r=0)$, meaning that this ansatz has an extra condition on possible solutions. However, with the translation symmetry this requirement can always be satisfied for $B\neq 0$.} At $B=1$ and $2$, where $p<3$, it decays faster than the
background tail and leaves the asymptotics of $q_{\rm base}$ unchanged. The
boundary conditions are therefore those of $q_{\rm base}$ exactly. $B$ is inherited as long as
$\|g\,\mathbf{v}\|<1$ pointwise, which keeps $q$ a continuous deformation of
$q_{\rm base}$.
However, the condition is not enforced during training. We monitor $B$
instead. In every anchored run reported in appendix~\ref{app:charge}, the nearest integer to $B$ is the target charge, and the appendix gives the deviations $|B-n|$. The last layer of the
network is initialised to zero, so
training starts from the ansatz itself, with $E_0=2.41627$ for $B=2$. This
track measures how much of the $2.5\%$ gap between the ansatz and the
reference energy~\cite{battye2002,battye2001} the network closes.

The two tracks answer different questions. The anchored track fixes the
topology and the boundary conditions, so the runs keep their topological
charge; it is the setting for the controlled
comparison of section~\ref{sec:fair}. The free track imposes nothing, so
runs can fail; it is the setting for the ansatz-free solutions of
section~\ref{sec:free}, and the comparison is repeated there under a single
protocol, with one set of hyperparameters used for all architectures and
all charges (section~\ref{sec:protocol}). The two tracks use different representations
and different initial data, and agree on the energy at $B=2$ to
$1.2\times10^{-5}$ on a common evaluation grid.

\subsection{Loss and randomised quadrature}
\label{sec:loss}
The loss is the energy itself plus a charge penalty,
\begin{equation}
L \;=\; E[q] \;+\; w_{\rm topo}\,\big(B[q]-B_{\rm t}\big)^2,
\label{eq:loss}
\end{equation}
where $B_{\rm t}$ is the target charge, with $w_{\rm topo}=200$ in the
anchored track and $1000$ in the free track.
In the anchored track the field inherits the charge of $q_{\rm base}$ as
long as $\|g\,\mathbf{v}\|<1$ (section~\ref{sec:field}), so the
penalty is inert up to quadrature error and serves only as a guard. In the free track the penalty acts mainly during the first epochs, while the far field
settles to the vacuum (section~\ref{sec:imprint}). The free track also uses a weak vacuum term
$\tfrac{1}{4}\,w_{\rm far}\langle\|q-q_{\rm vac}\|^2\rangle$, with $w_{\rm far}=50$,
evaluated on random points with $r\in[30,60]$. This
term induces the boundary condition that the anchored track has by
construction. Stationary points of $E$ solve the field equations, so no
second-derivative PDE residual is needed. Both the energy density and the
baryon density are assembled
from the same first-derivative Jacobian $\partial_iq$, obtained by
automatic differentiation, as in equation~\eqref{eq:densities}.

The energy integral runs over all of $\mathbb{R}^3$. To apply a quadrature rule on a finite interval, we replace the radial variable $r$ by $t$,
where $r=r_0t/(1-t)$. The map $t\mapsto r$ takes
$t\in[0,1)$ to $r\in[0,\infty)$, with $t=\tfrac12$ at $r=r_0$, so that points
spaced evenly in $t$ are dense in $r$ near the origin and spread out in the
tail. The quadrature is a tensor product in
$(t,\cos\theta,\phi)$: Gauss--Legendre in $t$ and in $\cos\theta$, and the
uniform rule in $\phi$, which converges exponentially for analytic
periodic integrands. Each node $(t,\cos\theta,\phi)$ gives a point $\mathbf{x}\in\mathbb{R}^3$ through
$r=r_0t/(1-t)$ and the angles $\theta$ and $\phi$, and the network receives
$\mathbf{u}=\mathbf{x}/(1+r)$ at that point (section~\ref{sec:field}).
The derivatives $\partial_iq$ are taken with respect to $\mathbf{x}$.
Densities are computed in
single precision and the quadrature sums are accumulated in double
precision, so that the many additions do not compound the single-precision
error. The integral is truncated at $r_{\max}=60$, that is at
$t=r_{\max}/(r_0+r_{\max})$. We place the cutoff this far out because a
cutoff at too small a radius omits part of the energy and lowers the
measured value, and the calculation itself gives no sign of it. The $B=1$
hedgehog profile used for the anchored background makes the omission
measurable. Its energy is $1.231446$ when the radial integral is carried to
infinity. This is the value quoted in section~\ref{sec:findings}, to six
digits. On $r\le60$ the same profile gives $1.231442$. The difference,
$4\times10^{-6}$, is the energy beyond the cutoff. At $B=2$ the profile
decays faster and the same cutoff omits less than $4\times10^{-7}$.

The training grid has $32\times20\times24$ points in
$(t,\cos\theta,\phi)$. The loss is the quadrature sum, not the integral itself, and on a
fixed set of nodes a network can lower the sum while the energy between the
nodes stays high~\cite{rivera2022}. We therefore randomise the grid at
every training step, by a fresh $SO(3)$ rotation of the whole grid and a new
value of $r_0$ drawn from $\{2.4,2.7,3.0,3.3,3.6\}$. The rotation moves the
nodes in angle, and, since it preserves $|\mathbf{x}|$, only the new $r_0$
moves them in radius. The same node positions therefore never recur.

In both tracks the energy minimisation uses Adam~\cite{kingma2015} with a
cosine learning-rate schedule~\cite{loshchilov2017} that decays to $2\%$ of
the initial rate.

\subsection{Architectures and implementations}
\label{sec:arch}
We compare MLPs and KANs. The MLPs are fully connected, with widths from
$2$ to $64$, depths from $1$ to $4$, and tanh or
GELU~\cite{hendrycks2016} activations. Each KAN layer maps an input vector
$u\in\mathbb{R}^{n_{\rm in}}$ to an output vector
$y\in\mathbb{R}^{n_{\rm out}}$ as the sum of two branches. The first applies
a fixed activation and then a trainable linear map, as an MLP layer does.
The second replaces the single weight of each connection by a spline in
$u_i$,
\begin{equation}
y_o=\sum_{i=1}^{n_{\rm in}} W_{oi}\,\mathrm{silu}(u_i)
+\sum_{i=1}^{n_{\rm in}}\sum_{m=1}^{G+k} c_{oim}\,N_m(u_i),
\qquad o=1,\dots,n_{\rm out}.
\end{equation}
Here $\mathrm{silu}(z)=z/(1+e^{-z})$, the $N_m$ are the B-splines of degree
$k=3$ on a uniform grid of $G=5$ intervals over $[-1,1]$, and the
weights $W_{oi}$ and spline coefficients $c_{oim}$ are trainable,
so each connection carries $G+k$ spline coefficients where an MLP would
have one weight. The symbols $u$ and
$y$ are local to this subsection: $u$ is the input of the layer in
question, and equals the compactified coordinate $\mathbf{u}$ only for the
first layer. We write $[w]{\times}d$ for a network of width $w$ and depth
$d$. The depth $d$ is the number of hidden layers, and $w$ is the width of
each hidden layer. For example, KAN$[16]{\times}4$ has four hidden layers
of width $16$, so its widths are $3$, $16$, $16$, $16$, $16$ and $4$ from
the input to the output.

In the official implementation \texttt{pykan}
(v0.2.8)~\cite{liu2024kan} the CPU spends most of a training step sending
operations to the GPU, so we wrote our own implementation of the same
model, with
batched Cox--de~Boor spline evaluation in the style of
\texttt{efficient-kan}~\cite{effkan}. All KAN results below use our
implementation. Appendix~\ref{app:implcost} compares the speed of our
implementation with that of \texttt{pykan}. We also ran
\texttt{FastKAN}~\cite{li2024fastkan}, a variant that replaces the splines
by radial basis functions. No result of the main text uses a \texttt{FastKAN}
run, and appendices~\ref{app:charge} and~\ref{app:quadfail} report these runs.

\subsection{Imprinting initial data (free track)}
\label{sec:imprint}
In the free track the topological sector is selected by \emph{imprinting}.
A crude initial configuration with the desired topological charge is built as a
product ansatz of $B$ unit-charge hedgehog lumps,
\begin{equation}
q_{\rm init}=\prod_{j=1}^{B} q^{(j)},\qquad
q^{(j)}=\big(\cos f_{\rm lump}(\rho_j),\,\sin f_{\rm lump}(\rho_j)\,
\mathcal{R}_j\hat{\boldsymbol\rho}_j\big),
\qquad f_{\rm lump}(\rho)=\pi e^{-\rho}.
\label{eq:lumps}
\end{equation}
Here $\rho_j=|\mathbf{x}-\mathbf{c}_j|$ and
$\hat{\boldsymbol\rho}_j=(\mathbf{x}-\mathbf{c}_j)/\rho_j$ are measured from
the centre $\mathbf{c}_j$ of the $j$th lump, and the product
is taken in SU(2), that is, by quaternion multiplication. The centres $\mathbf{c}_j$ are drawn
from an isotropic Gaussian of standard deviation $1.5$, shifted so that
their mean is at the origin, and rescaled if necessary so that
$\max_j|\mathbf{c}_j|\le2.5$, because the map $\mathbf{x}\mapsto\mathbf{u}$
takes lumps far from the origin into a thin layer next to the boundary
$|\mathbf{u}|=1$, where $q_{\rm init}$ varies too rapidly for the network to
fit. The
$\mathcal{R}_j\in SO(3)$ are independent random isospin orientations. This initial state carries the
topological charge but no information about the target symmetry,\footnote{At
$B=1$ the initial state is a single lump at the origin, which is already
spherically symmetric.} no optimised profile
and no tuning of the relative orientations, which are random rather than
chosen so that the lumps attract. The network is fitted to
$q_{\rm init}$ by plain mean-square regression and is then released into
energy minimisation. The regression is applied to the raw output
$\mathbf{v}$ before normalisation. Since $\|q_{\rm init}\|=1$, fitting
$\mathbf{v}$ to $q_{\rm init}$ also brings $q$ to it. We fit $\mathbf{v}$ rather than $q$ because $q=\mathbf{v}/\|\mathbf{v}\|$ does not depend on the length of $\mathbf{v}$, and $q$ is not defined where $\mathbf{v}=0$. In the uniform protocol it uses $8000$ Adam steps at a constant learning rate of $10^{-3}$ on batches
of points sampled uniformly in a ball of radius $15$.\footnote{Each batch
draws $2048$ points uniformly in the cube $[-15,15]^3$ and keeps the points
inside the ball, on average $2048\pi/6\approx1072$ points.} This
regression loss, which we call the imprinting loss, is the quantity analysed
in section~\ref{sec:freeze}.

In the free track the role of the charge penalty of equation~\eqref{eq:loss} is
transient. During the first epochs after imprinting the far field has not yet
settled to the vacuum: the imprinting regression samples only $r\le15$, while
$B$ is measured out to $r_{\max}$ (section~\ref{sec:loss}). The field is then
not a map of the compactified domain, so $B$ is not quantised, the value
measured on the quadrature grid can move continuously, and the charge penalty acts on that value. Once $q\approx q_{\rm vac}$ in the far region, $B$ is the degree of a map
$S^3\to S^3$ and is invariant under continuous deformation of the field
(section~\ref{sec:model}); a training step is such a deformation except where
the field becomes singular, so the penalty supplies almost no gradient and
becomes inert (section~\ref{sec:freeze}).

\subsection{Evaluation, verification and success}
\label{sec:eval}
After training, the final energy $E$, the charge $B$ and the virial ratio $E_2/E_4$ are computed again on an evaluation grid
of $48\times28\times36$ points. We evaluate them three times, with
$r_0=2.6$, $3.0$ and $3.4$ and a random rotation for each. The reported $E$
and $B$ are the means of the three values, and the reported $E_2/E_4$ is the
ratio of the mean $E_2$ to the mean $E_4$. For every run we also report the
spread $\Delta E\equiv E_{\max}-E_{\min}$ of $E$ across these three
evaluations. All the grids compared by $\Delta E$ have the
same number of points, so $\Delta E$ does not show how $E$ changes when
more points are used.

A run is judged at two levels: verification and success.

We apply the verification protocol to every run that this paper reports.
The verification protocol uses two kinds of check. The first kind tests exact properties of the Skyrme model: (i) the deviation $|B-n|$
from the nearest integer $n$, (ii) the virial ratio $E_2/E_4$, and (iii) the
Faddeev--Bogomolny bound~\eqref{eq:fb}. The second kind tests the accuracy of the numerical integration: (iv) the spread $\Delta E$ of $E$ across the evaluation grids described above. The thresholds of checks (i), (ii)
and (iv) are $|B-n|<10^{-3}$, $|E_2/E_4-1|<0.05$ and $\Delta E<10^{-2}$.
Check (iii) requires $E\ge|B|$. We emphasize that none of
these checks uses a reference solution.

Success is defined only for the free track. We say that a run
succeeds if it passes the verification protocol
with $B$ at the target charge, and its energy is within $1\%$ of the
published reference energy of table~\ref{tab:bounds}~\cite{battye2002,battye2001}.\footnote{No run of this paper that passes the verification protocol with $B$ at the target charge has an energy more than $1\%$ below the reference energy.} To see how $E$ changes when
more points are used, we also compute $E$ in the same way on a grid of
$80\times44\times60$ points
for every successful run of section~\ref{sec:free}. Section~\ref{sec:findings}
reports the change from the $48\times28\times36$ grid. Table~\ref{tab:failures} and appendix~\ref{app:modes} sort the unsuccessful runs of table~\ref{tab:uniform} into three failure modes.

\subsection{Experimental protocol}
\label{sec:protocol}
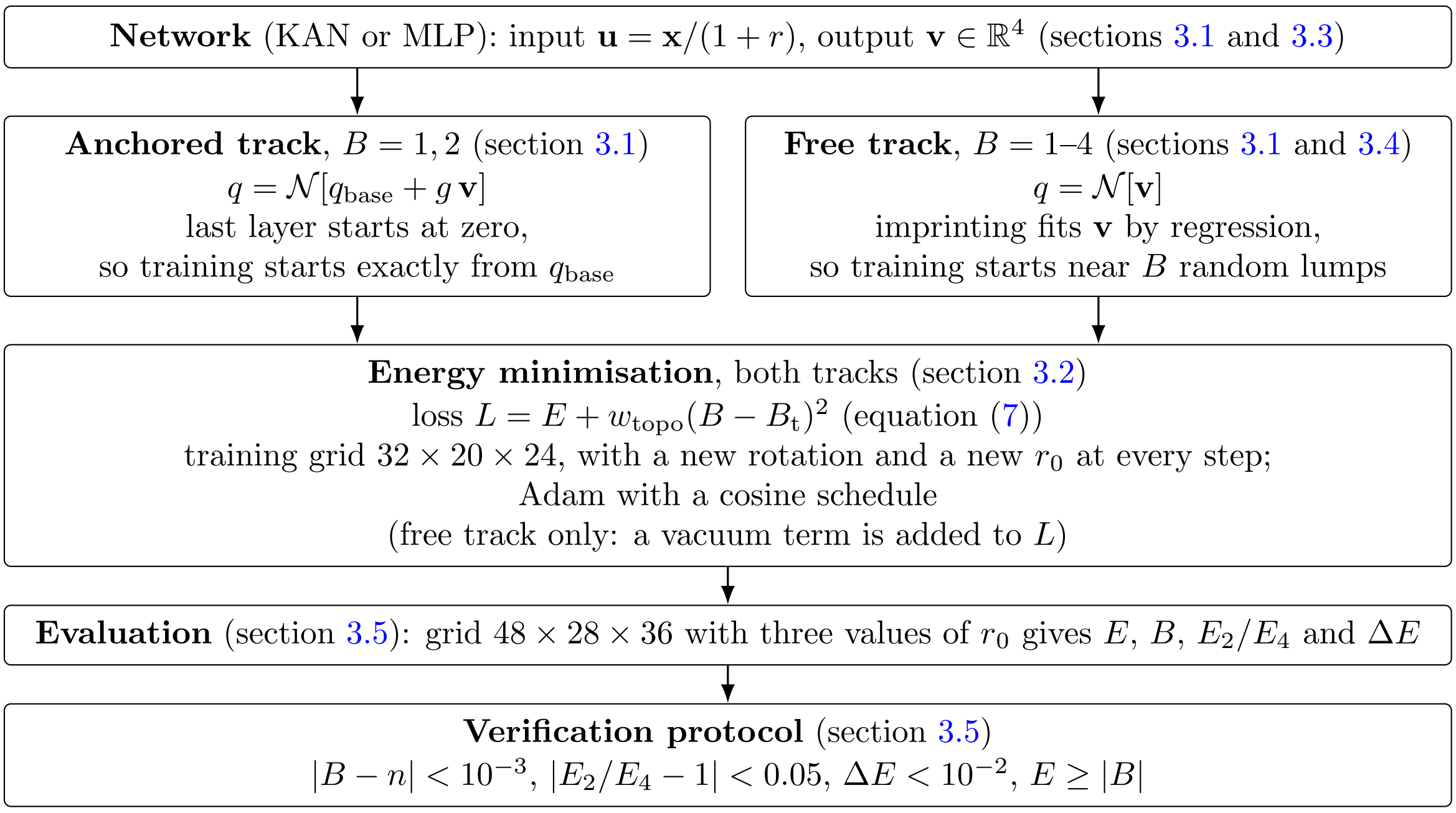
\begin{figure}[t]
\centering
\begin{tikzpicture}[
  font=\small,
  box/.style={draw, rounded corners=2pt, align=flush center, inner sep=4pt},
  arr/.style={-{Latex[length=2mm]}, semithick},
  node distance=4mm and 4mm]
\node[box, text width=0.92\textwidth] (net)
  {\textbf{Network} (KAN or MLP): input $\mathbf{u}=\mathbf{x}/(1+r)$, output
   $\mathbf{v}\in\mathbb{R}^4$ (sections~\ref{sec:field} and~\ref{sec:arch})};
\node[box, text width=0.44\textwidth, anchor=north west]
  (anc) at ($(net.south west)+(0,-5mm)$)
  {\textbf{Anchored track}, $B=1,2$ (section~\ref{sec:field})\\[1pt]
   $q=\mathcal{N}[q_{\rm base}+g\,\mathbf{v}]$\\
   last layer starts at zero,\\ so training starts exactly from $q_{\rm base}$};
\node[box, text width=0.44\textwidth, anchor=north east]
  (free) at ($(net.south east)+(0,-5mm)$)
  {\textbf{Free track}, $B=1$--$4$ (sections~\ref{sec:field} and~\ref{sec:imprint})\\[1pt]
   $q=\mathcal{N}[\mathbf{v}]$\\
   imprinting fits $\mathbf{v}$ by regression,\\ so training starts near $B$ random lumps};
\node[box, text width=0.92\textwidth, anchor=north]
  (train) at ($(net.south |- free.south)+(0,-5mm)$)
  {\textbf{Energy minimisation}, both tracks (section~\ref{sec:loss})\\[1pt]
   loss $L=E+w_{\rm topo}(B-B_{\rm t})^2$ (equation~\eqref{eq:loss})\\
   training grid $32\times20\times24$, with a new rotation and a new $r_0$ at
   every step;\\ Adam with a cosine schedule\\
   (free track only: a vacuum term is added to $L$)};
\node[box, text width=0.92\textwidth, below=of train] (eval)
  {\textbf{Evaluation} (section~\ref{sec:eval}): grid $48\times28\times36$
   with three values of $r_0$ gives $E$, $B$, $E_2/E_4$ and $\Delta E$};
\node[box, text width=0.92\textwidth, below=of eval] (ver)
  {\textbf{Verification protocol} (section~\ref{sec:eval})\\[1pt]
   $|B-n|<10^{-3}$, $|E_2/E_4-1|<0.05$, $\Delta E<10^{-2}$, $E\ge|B|$};
\draw[arr] (net.south -| anc.north) -- (anc.north);
\draw[arr] (net.south -| free.north) -- (free.north);
\draw[arr] (anc.south) -- (anc.south |- train.north);
\draw[arr] (free.south) -- (free.south |- train.north);
\draw[arr] (train) -- (eval);
\draw[arr] (eval) -- (ver);
\end{tikzpicture}
\caption{Steps of a run, from the network to the verification protocol.
The box on the left is for the anchored track, and the box on the right is
for the free track. Boxes across the full width are for both tracks. Each
box starts with its name in bold and gives the section that describes the
step.}
\label{fig:flow}
\end{figure}
Figure~\ref{fig:flow} shows the steps of a run in the two tracks.
Every run fixes its random seed and stores its settings --- hyperparameters,
which of the two tracks it uses, the target charge and the quadrature grids ---
together with its training history and verification metrics, on disk. A
re-run of a run is a new run that uses the same settings and the same seed
as that run. Unless stated
otherwise, the anchored track uses the same settings for every model:
$1000$ epochs of Adam with a cosine schedule and a learning rate of
$2\times10^{-3}$. Section~\ref{sec:fair} examines what changes when the
learning rate is tuned for each model. The free
track uses a uniform protocol, with the same hyperparameters for every network
and every charge: $8000$ imprinting steps of Adam at a constant learning rate of
$10^{-3}$ (section~\ref{sec:imprint}), and then $20\,000$ energy-minimisation
epochs of Adam with a cosine schedule and a learning rate of $2\times10^{-3}$. All experiments ran on a single NVIDIA RTX 6000
Ada GPU (48\,GB) in PyTorch. A $1000$-epoch anchored run takes $20$ to
$24\,$s for an MLP and $30$ to $120\,$s for a KAN, and a full free-track run
with imprinting and $20\,000$ epochs takes
$3$ to $12\,$min depending on the architecture.

\section{Results I: controlled comparison on the $B=2$ benchmark}
\label{sec:fair}

\begin{figure}[t]
\centering
\includegraphics[width=\textwidth]{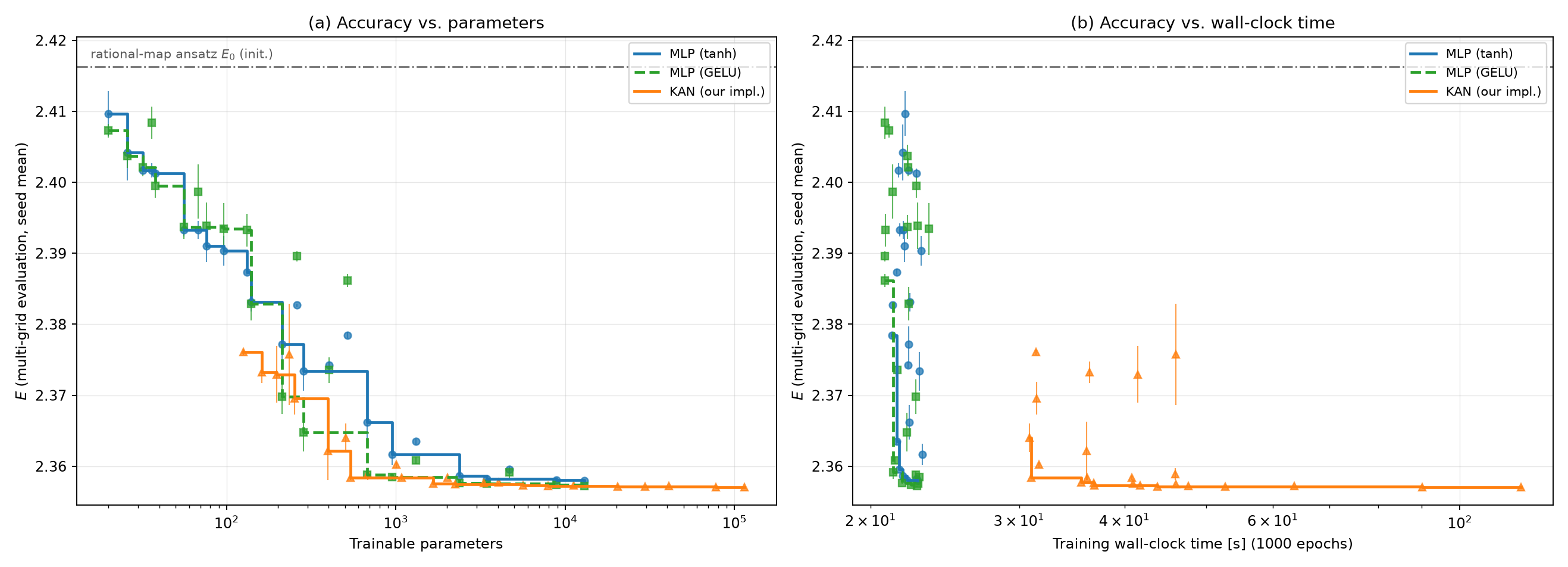}
\caption{Anchored-track comparison at $B=2$. Each point is one
(width, depth) pair, averaged over five seeds; error bars are seed standard
deviations. The median of these deviations over the $24$ pairs is
$1.0\times10^{-4}$ for the KAN, $7.7\times10^{-4}$ for the tanh-MLP and
$1.5\times10^{-3}$ for the GELU-MLP. Each step line is a Pareto front: in (a) the lowest energy
reached using no more parameters than that, and in (b) the lowest reached in
no more time. The horizontal line is $E_0=2.41627$, the energy of the
rational-map ansatz, where every run in this figure starts.}
\label{fig:pareto}
\end{figure}

We define the anchored sweep as the $600$ anchored runs at the six widths
$2$, $4$, $8$, $16$, $32$ and $64$ and the four depths $1$ to $4$, $24$
(width, depth) pairs in all, at five seeds each. It contains the KAN, the
tanh-MLP and the GELU-MLP at $B=2$, and the KAN and the tanh-MLP at $B=1$.
The following settings are the same for every architecture: the loss of
equation~\eqref{eq:loss} with $w_{\rm topo}=200$, the quadrature of
section~\ref{sec:loss} on the training grid of $32\times20\times24$ points,
$1000$ epochs of Adam with a cosine schedule and a learning rate of
$2\times10^{-3}$ (section~\ref{sec:protocol}), and the evaluation grid of
$48\times28\times36$ points (section~\ref{sec:eval}). Figure~\ref{fig:pareto} summarises the $360$ runs at
$B=2$, and appendix~\ref{app:charge} uses the $240$ runs at $B=1$. Every
run passes the verification protocol of section~\ref{sec:eval}.

\begin{table}[t]
\centering
\caption{Fairness-controlled comparison at $B=2$. Each entry is a mean
energy over five seeds, with its standard error in parentheses in units of
the last digit. The (width, depth) column says where the pair of each
row comes from. In the first row each entry comes from the pair whose mean
is the lowest among the $24$ pairs of the anchored sweep; this pair is
$[64]{\times}4$ for all three networks. In rows 2 to 4, the fixed pair is
$[32]{\times}3$ for the KAN and $[64]{\times}4$ for both MLPs. The second
row takes the lowest mean among the learning rates $5\times10^{-4}$,
$10^{-3}$, $2\times10^{-3}$, $5\times10^{-3}$ and $10^{-2}$. In the third
and fourth rows the learning rate is $2\times10^{-3}$, and the training
stops after $30\,$s and after $120\,$s. Bold marks the lowest entry of a
row when it is below the next by more than three standard errors of the
difference. All entries are evaluated on the $48\times28\times36$ grid.
The published grid value at $B=2$ is $2.3582$ (table~\ref{tab:bounds}).}
\begin{tabular}{lllll}
\toprule
Comparison axis & (width, depth) & KAN & tanh-MLP & GELU-MLP \\
\midrule
Shared learning rate ($2\times10^{-3}$) & lowest of $24$ & $\mathbf{2.357053(22)}$ & $2.357939(17)$ & $2.357197(25)$ \\
\midrule
Learning rate tuned per model & fixed & $\mathbf{2.356997(5)}$ & $2.357427(39)$ & $2.357048(12)$ \\
Within $30\,$s & fixed & $2.357219(15)$ & $2.357811(18)$ & $\mathbf{2.357120(18)}$ \\
Within $120\,$s & fixed & $2.357028(14)$ & $2.357113(3)$ & $2.356989(3)$ \\
\bottomrule
\end{tabular}
\label{tab:fair}
\end{table}

Three findings stand out (table~\ref{tab:fair} and figure~\ref{fig:pareto}).
First, with the untuned
defaults --- the tanh activation for the MLP, and the learning rate of
$2\times10^{-3}$ used for every model --- the KAN dominates. Its lowest mean energy among the $24$ (width, depth) pairs is below that of the tanh-MLP (first row of table~\ref{tab:fair}), and it needs fewer
parameters than the tanh-MLP at matched accuracy
(figure~\ref{fig:pareto}(a)). The KAN's energies also
vary less from seed to seed over most of the sweep. Second, and centrally, these two gaps in mean energy and in the number of parameters are
largely a property of those defaults and not of the architecture. Replacing tanh by GELU alone brings the lowest mean energy of the MLP from
$2.35794$ to $2.35720$. The remaining differences are real but modest.
In figure~\ref{fig:pareto}(a), the KAN front is lower than the GELU-MLP front between about
$4\times10^2$ and $10^4$ parameters. The parameter ratio of the GELU-MLP to the KAN at matched
accuracy varies from $0.6$ to $4.4$ with the energy: at some energies the
KAN needs more parameters, not fewer.
Third, which model is ahead depends on what is held fixed. Rows 2 to 4 of
table~\ref{tab:fair} use one (width, depth) pair of each network,
$[32]{\times}3$ for the KAN and $[64]{\times}4$ for both MLPs.\footnote{These
pairs are not matched in the number of parameters, and matching them could
change the differences in rows 2 to 4. Appendix~\ref{app:fair} gives two
checks. For rows~3 and~4, it compares a KAN with a tanh-MLP that has about
the same number of parameters. For row~2, it compares a KAN with the two
MLPs of the table, which have more parameters than this KAN.} When the
learning rate is tuned for each model, the KAN, which has $1.6$ times as
many parameters as the GELU-MLP, is still lower than the GELU-MLP at a
matched number of epochs, by $5\times10^{-5}$. A KAN epoch costs more --- $31$ to $118\,$s per thousand
against $21$ to $23\,$s for either MLP (figure~\ref{fig:pareto}(b)) --- so at
a matched time budget the ordering changes: at the shared learning rate, within $30\,$s the GELU-MLP is
lower by $1\times10^{-4}$, and within $120\,$s the two are not separated.
With the learning rate tuned for each model (row~2), the tanh-MLP is
higher than the KAN and the GELU-MLP by about $4\times10^{-4}$. For all
three networks, the lowest mean energy is at a learning rate of $10^{-2}$,
the largest rate that we tried.\footnote{For the GELU-MLP, the mean energy
at a learning rate of $5\times10^{-3}$ is higher by $4\times10^{-5}$, or $1.9$ standard errors of
the difference, so the two rates are not separated by the
three-standard-error criterion of table~\ref{tab:fair}. For the KAN, the
mean energy at $5\times10^{-3}$ is higher by $3\times10^{-5}$, or $3.0$
standard errors, close to the criterion. With five more seeds of the KAN
at each of the two rates ($10$ runs), the difference over ten seeds is
$4.8$ standard errors.}
Appendix~\ref{app:fairapp} describes the fairness checklist, covering
learning rate, activation,
implementation and wall-clock budget.

We tested whether the entries of table~\ref{tab:fair} depend on the
evaluation grid as follows. For each of the $12$ entries, we re-ran
the five runs (one per seed) whose mean energy is the entry, $60$ runs in
all.\footnote{For the KAN entry of the first
row, the re-runs use KAN$[64]{\times}3$ instead of KAN$[64]{\times}4$. The
re-runs were made when table~\ref{tab:fair} used three seeds, and with three
seeds KAN$[64]{\times}3$ had the lowest mean energy. With five seeds
KAN$[64]{\times}4$ is lower, by $8\times10^{-7}$.} Each re-run uses the same
network, (width, depth) pair, learning rate, seed, and number of epochs or
time limit as the run that it repeats. We saved the models of the re-runs
and evaluated them on the finer $80\times44\times60$ grid as well. For each
saved model, we compared the energy on the $80\times44\times60$ grid with
the energy on the $48\times28\times36$ grid. The largest difference is
$3.3\times10^{-5}$ for the KAN and $2.4\times10^{-7}$ for the MLPs. Both are
smaller than every difference that table~\ref{tab:fair} marks in bold.

\section{Results II: ansatz-free solutions and spontaneous symmetry formation}
\label{sec:free}

\subsection{Uniform protocol and success rates}
\label{sec:uniform}
Free-track runs are initialised by imprinting random multi-lump data
(section~\ref{sec:imprint}) and then relaxed by energy minimisation. Section~\ref{sec:eval} defines when a run succeeds.

Table~\ref{tab:uniform} reports the uniform-protocol study, which covers
charges $B=1$ to $4$. It compares a KAN and a GELU-MLP at each of two
parameter counts, about $8\times10^3$ and about $1.3\times10^4$:
KAN$[16]{\times}4$ with GELU-MLP$[48]{\times}4$, and KAN$[21]{\times}4$ with
GELU-MLP$[64]{\times}4$.\footnote{The (width, depth) pairs of
KAN$[16]{\times}4$ and GELU-MLP$[64]{\times}4$ are two of the $24$ pairs of
the anchored sweep (section~\ref{sec:fair}).} Each cell of table~\ref{tab:uniform} gives the result of
one network at one charge. Each network is run ten times at $B=1$ and at
$B=2$, and twenty times at $B\ge3$. The $B\ge3$ cells were extended from
ten runs for statistical power after the first ten had
been seen, with the protocol unchanged and every run reported
(section~\ref{sec:limitations}). The runs of one cell differ
only in the random seed, which fixes the lump centres and orientations, the
network initialisation and the grid rotations used during training. At
$B=1$ the single lump is always at the origin, so the seed changes the
orientation of the lump but not its position.
Apart from the network, all hyperparameters are identical across cells: the
learning rate, the epoch budget, the imprinting stage and the two penalty
weights $w_{\rm topo}$ and $w_{\rm far}$ of section~\ref{sec:loss}.

At $B=1$ and $B=2$ almost every run succeeds: seven of
the eight cells are $10/10$, and the eighth is $9/10$. That single failure
meets every condition except the energy: it stops $1.9\%$ above the
reference energy~\cite{battye2002,battye2001}. The networks
separate only at $B\ge3$, and there
the success rates follow the pattern of section~\ref{sec:fair}. At equal epoch
budget and about $8\times10^3$ parameters KAN$[16]{\times}4$ succeeds more
often than GELU-MLP$[48]{\times}4$: at $B=3$ the rates
are $19/20$ against $13/20$, with a two-sided Fisher exact $p=0.044$;
at $B=4$ they are $11/20$ against $6/20$, with $p=0.20$.
These $p$-values are not corrected for the number of tests
(section~\ref{sec:limitations}). At about
$1.3\times10^4$ parameters the two are indistinguishable:
KAN$[21]{\times}4$ against GELU-MLP$[64]{\times}4$ gives $17/20$ against
$16/20$ at $B=3$ ($p=1.0$) and $12/20$ against $10/20$ at $B=4$ ($p=0.75$). The two KAN sizes are
indistinguishable as well: KAN$[16]{\times}4$ against KAN$[21]{\times}4$ gives
$19/20$ against $17/20$ at $B=3$ ($p=0.60$) and $11/20$ against $12/20$ at
$B=4$ ($p=1.0$). In other words,
the MLP reaches the success rate of the KAN by adding parameters, while the
KAN shows no further gain beyond $8\times10^3$ parameters. This mirrors
figure~\ref{fig:pareto}(a), where the KAN front is lower between about
$4\times10^2$ and $10^4$ parameters: $8\times10^3$ lies inside that range and
$1.3\times10^4$ outside it.\footnote{The gap is too small to read off the figure. At $8\times10^3$ parameters the fronts are at $2.35721$ (KAN) and $2.35751$ (GELU-MLP); at $1.3\times10^4$ they are at $2.35721$ and $2.35720$.}

\begin{figure}[!t]
\centering
\includegraphics[width=0.92\textwidth]{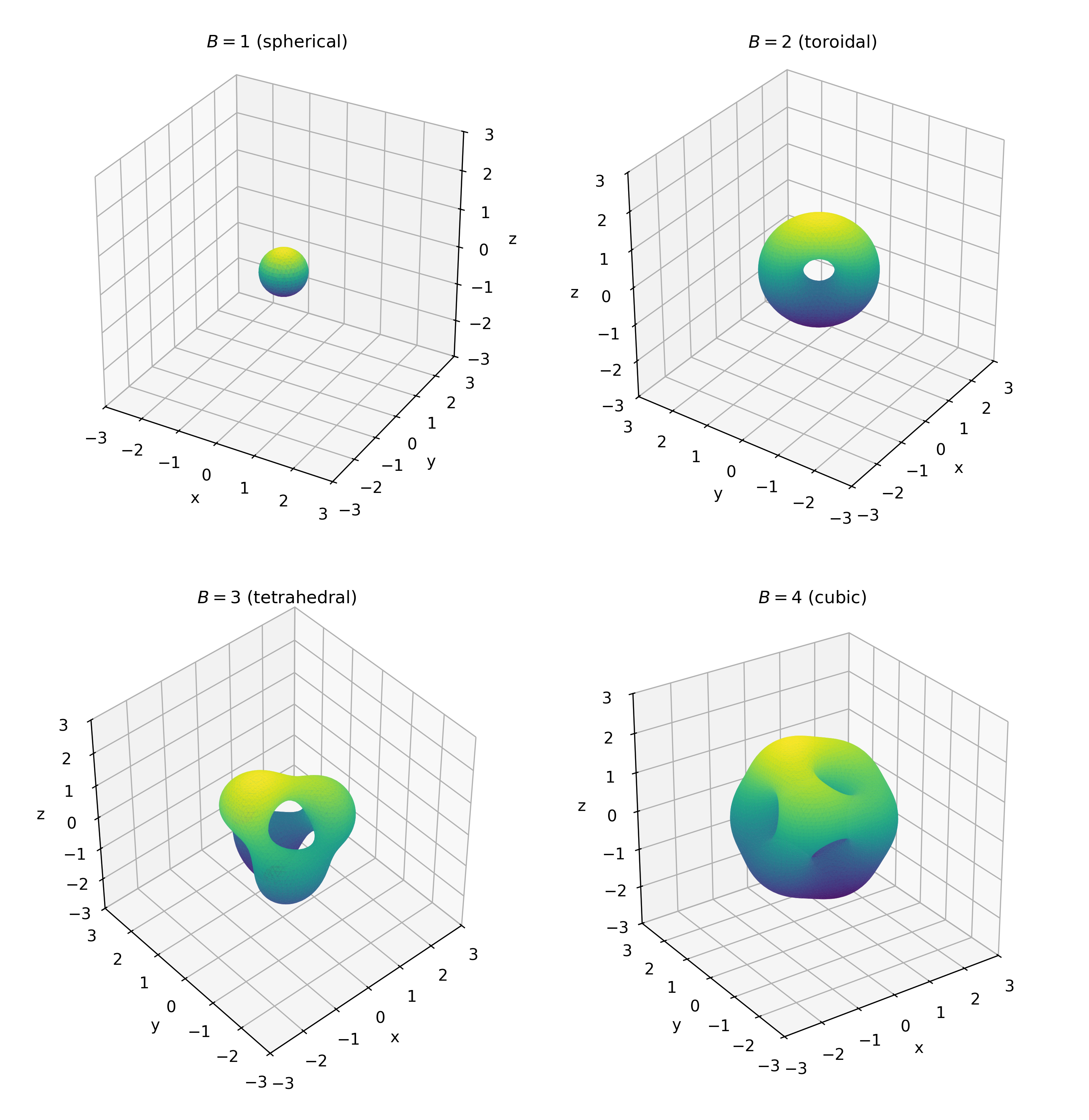}
\caption{Baryon-density isosurfaces (at $50\%$ of maximum) of the four
headline solutions (section~\ref{sec:uniform}), which are free-track runs.
Torus, tetrahedron and cube emerge from random
multi-lump initial data that contains no symmetry information. At $B=1$ the
initial data is a single lump at the origin, which is already spherically
symmetric. The shapes
are identified quantitatively in the text, not from these images. Colour
indicates height along $z$ and carries no physical information.}
\label{fig:iso}
\end{figure}

Among the successful runs of section~\ref{sec:free}, we take at each charge
the run with the lowest energy on the $80\times44\times60$ grid. We call
this run the headline solution of that charge. The headline solutions were
obtained with GELU-MLP$[48]{\times}4$ at $B=1$ and with GELU-MLP$[64]{\times}4$ at
$B=2$--$4$. Those at $B=1$ and $B=2$ come from table~\ref{tab:uniform}, and
those at $B=3$ and $B=4$ come from the $660\,$s runs of table~\ref{tab:wallclock}.
Table~\ref{tab:headline} in appendix~\ref{app:headline} gives the details of
each headline solution. Figure~\ref{fig:iso} and table~\ref{tab:bounds} show the four headline solutions; section~\ref{sec:symmetry} analyses their symmetry and section~\ref{sec:findings} their energies.

\begin{table}[t]
\centering
\caption{Uniform-protocol free-track study. Each entry is the number of
successes out of the number of runs. There are ten runs at $B\le2$ and
twenty at $B\ge3$. The trainable parameter count is given under each network
name.}
\begin{tabular}{ccccc}
\toprule
$B$ & KAN $[16]{\times}4$ & KAN $[21]{\times}4$ & GELU-MLP $[64]{\times}4$ &
GELU-MLP $[48]{\times}4$ \\
 & (7\,920 par.) & (13\,230 par.) & (12\,996 par.) & (7\,444 par.) \\
\midrule
1 & 10/10 & 10/10 & 10/10 & 10/10 \\
2 & 10/10 & 10/10 & 10/10 & 9/10 \\
3 & 19/20 & 17/20 & 16/20 & 13/20 \\
4 & 11/20 & 12/20 & 10/20 & 6/20 \\
\bottomrule
\end{tabular}
\label{tab:uniform}
\end{table}

\subsection{Symmetry of the solutions}
\label{sec:symmetry}
Figure~\ref{fig:iso} shows the baryon-density isosurfaces of the four headline solutions. At $B=2$--$4$ the runs start from lumps at random positions and orientations (section~\ref{sec:imprint}), and the torus, the tetrahedron and the cube known for these charges are visible at once. The eye can recognise these shapes but cannot tell how exactly the symmetry holds, so we quantify it: we compute the inertia tensor weighted by the energy
density~\eqref{eq:densities},
\begin{equation}
T_{ij}=\int e(\mathbf{x})\,
\big(\delta_{ij}\,|\tilde{\mathbf{x}}|^2-\tilde{x}_i\tilde{x}_j\big)\,
\mathrm{d}^3x,
\qquad \tilde{\mathbf{x}}=\mathbf{x}-\mathbf{x}_{\rm c},
\label{eq:inertia}
\end{equation}
where $\delta_{ij}$ is the Kronecker delta,
$\mathbf{x}_{\rm c}=\int e(\mathbf{x})\,\mathbf{x}\,\mathrm{d}^3x
\big/\int e(\mathbf{x})\,\mathrm{d}^3x$ is the energy centroid, evaluated on
a $48\times28\times36$ grid with $r_0=3$, truncated at $r=30$.\footnote{We truncate the integral at $r=30$
because the weight $|\tilde{\mathbf{x}}|^2$ makes $T_{ij}$ sensitive to the
far field. With cutoffs from $15$ to $40$, the four values in the last
column of table~\ref{tab:headline} stay between $7\times10^{-5}$ and
$1.5\times10^{-4}$.} The three eigenvalues of $T_{ij}$ are
the principal moments. We order the principal moments as
$I_1\le I_2\le I_3$. Each symmetry class requires some of the three to be
equal. A solution with spherical, tetrahedral ($T_d$) or cubic ($O_h$)
symmetry must have all three equal: none of these symmetry groups leaves a
traceless symmetric tensor invariant, so $T_{ij}$ can only be proportional
to $\delta_{ij}$. A solution with the axial symmetry of a torus
($D_{\infty h}$) must have only two of them equal, namely the
moments about the two axes perpendicular to the symmetry axis. The symmetry
group does not say whether the moment about the symmetry axis is the
largest or the smallest of the three. We therefore decide from the measured moments which two of the three form the equal pair. In the $B=2$ headline solution, the largest moment is
$23\%$ larger than each of the other two, so the equal pair is $I_1$ and
$I_2$.\footnote{The rational-map ansatz~\cite{manton2004,houghton1998}
with $R=z^2$ has the symmetry of the $B=2$ solution. We computed the principal moments of its
energy density: for any profile $f(r)$, the moment about the symmetry axis
is the largest of the three.} We measure how well each requirement is met by the relative
difference between the largest and the smallest of the moments that must
agree: $(I_3-I_1)/I_1$ for the spherical, tetrahedral and cubic cases, and
$(I_2-I_1)/I_1$ for the torus.
The headline solutions realise these degeneracies to between $8\times10^{-5}$ and $1.3\times10^{-4}$, without any
symmetry having been imposed.

Spherical, tetrahedral and cubic symmetry each require all three moments to be equal,
so the principal moments alone cannot distinguish them. We therefore also compute
the rotation-invariant multipole powers of the baryon
density~\eqref{eq:densities},
\begin{equation}
S_\ell=\Big(\sum_{m=-\ell}^{\ell}|Q_{\ell m}|^2\Big)^{1/2},
\qquad
Q_{\ell m}=\frac{\int b(\mathbf{x})\,r^\ell\,
Y_{\ell m}^{*}(\theta,\phi)\,\mathrm{d}^3x}
{\int b(\mathbf{x})\,r^\ell\,\mathrm{d}^3x},
\label{eq:multipole}
\end{equation}
which do not depend on the orientation of the solution. The radius $r$ and
the angles $(\theta,\phi)$ in equation~\eqref{eq:multipole} are measured from
the centroid of the baryon density, defined as in
equation~\eqref{eq:inertia} with $b$ in place of $e$. Here $Y_{\ell m}$ are
the spherical harmonics of degree $\ell$. Dividing by
$\int b\,r^\ell\,\mathrm{d}^3x$ makes $S_\ell$ dimensionless and independent
of the overall size of the solution, so $S_\ell$ measures the angular shape
of the baryon density at degree $\ell$ alone. Whether $S_\ell$ can be
nonzero is decided by the same argument that fixed the principal moments:
the symmetry group must leave some combination of the degree-$\ell$
spherical harmonics invariant.\footnote{The multipole analysis starts at
$\ell=2$: the dipole $S_1$ vanishes identically because $r$ is measured from
the centroid of the baryon density. Of the four symmetries, only that of the
torus leaves a combination of degree $2$ invariant. The degree-$3$
combination $xyz$ is invariant under tetrahedral symmetry, but the inversion
symmetry of $O_h$ and $D_{\infty h}$ forbids it, and spherical symmetry
forbids every combination of degree $3$. Every symmetry except the spherical
one leaves a combination of degree $4$ invariant, so $S_4$ is what separates
the cubic solution from the spherical one.} Each of the four symmetry classes gives a
different pattern of vanishing $S_\ell$, so the triple $(S_2,S_3,S_4)$
identifies the class uniquely. In table~\ref{tab:multipole} the entries that
must vanish lie at least two orders of magnitude below those that need not. These
numbers establish the symmetry assignment without any visual input.

\begin{table}[t]
\centering
\caption{Rotation-invariant multipole powers
(equation~\eqref{eq:multipole}) of the baryon density of the headline
solutions. Entries of order $10^{-5}$ are consistent with zero. Bold marks
the multipole that sets each class apart; the spherical solution has none,
because every $S_\ell$ vanishes for it. The fifth column names the
entries that symmetry requires to vanish, and the last column gives the
assignment.}
\begin{tabular}{lccccl}
\toprule
Solution & $S_2$ & $S_3$ & $S_4$ & must vanish & assignment \\
\midrule
$B{=}1$ & $1.8{\times}10^{-5}$ & $2.9{\times}10^{-5}$ & $5.0{\times}10^{-5}$
 & all three & spherical \\
$B{=}2$ & $\mathbf{1.6{\times}10^{-1}}$ & $1.5{\times}10^{-5}$
 & $2.7{\times}10^{-2}$ & $S_3$ & toroidal ($D_{\infty h}$) \\
$B{=}3$ & $4.3{\times}10^{-5}$ & $\mathbf{1.5{\times}10^{-1}}$
 & $8.6{\times}10^{-3}$ & $S_2$ & tetrahedral ($T_d$) \\
$B{=}4$ & $6.2{\times}10^{-5}$ & $6.3{\times}10^{-5}$
 & $\mathbf{1.3{\times}10^{-1}}$ & $S_2$ and $S_3$
 & cubic ($O_h$) \\
\bottomrule
\end{tabular}
\label{tab:multipole}
\end{table}

\subsection{Main findings}
\label{sec:findings}
Five points deserve emphasis. (i)~The known symmetries emerge on their own
(figure~\ref{fig:iso}). At $B=2$, for example, two lumps placed at a random
separation and orientation merge into a torus whose two smallest principal
moments agree to $8\times10^{-5}$. Appendix~\ref{app:relax} shows how the torus, the tetrahedron and the cube form from the lumps during training.

(ii)~The lowest energies reached at
$B=2$--$4$ lie below the published reference values~\cite{battye2002,battye2001} (table~\ref{tab:bounds})
by $0.052\%$, $0.045\%$ and $0.011\%$, respectively.\footnote{More recent high-resolution studies of the multi-Skyrmion landscape~\cite{feist2013,gudnason2022} include a pion mass term and are therefore not directly comparable.} To judge whether these three deficits are plausible, we turn to $B=1$, where the published value can be checked: the $B=1$ minimiser is believed, though not proven, to be spherically symmetric~\cite{manton2004}, and the spherically symmetric reduction is an ordinary differential
equation, which can be solved to arbitrary precision and gives $E=1.23145$.
The published value is $0.06\%$ above it. Whether
the same offset applies at higher topological charge is not known a priori. The published values at $B\ge2$, however, come from the same grid calculation as the published value at $B=1$~\cite{battye2002}, so an offset of similar size is plausible there. Our deficits at $B=2$--$4$ are all smaller than the $B=1$ offset of $0.06\%$. Moreover, among the free-track runs at $B\ge2$ that
pass the verification protocol with $B$ at the target charge, including
the headline solutions, none lies more than $0.06\%$ below the reference energy.
Every free-track run that falls more than $1\%$ below the reference energy ends
away from the target charge.

(iii)~The computed charge of our four solutions differs from $B$ by at most $10^{-6}$ (table~\ref{tab:bounds}). On the
grid that gave the four published values, by contrast, the computed
charge is $1.20$ to $1.51\%$ below $B$, more than four orders of magnitude further from $B$. Battye and Sutcliffe~\cite{battye2002,battye2001} correct for this by multiplying the raw grid energy by $B$ and dividing by the computed charge (table~\ref{tab:bounds}).
Our four energies need no such correction.

(iv)~The four headline solutions pass the verification protocol far inside its thresholds (appendix~\ref{app:headline}).
Two quantities of section~\ref{sec:eval} estimate the quadrature error on
the energies of these four solutions: the spread $\Delta E$ over independent grids of the
same resolution is at most $7\times10^{-7}$ on the $80\times44\times60$ grid
(appendix~\ref{app:headline}),
and the change of $E$ from the $48\times28\times36$ grid to the $80\times44\times60$ grid
is below that level. Our values are
variational upper bounds, so we offer them as
independent cross-checks of the reference energies~\cite{battye2002,battye2001}. The four solutions also satisfy the virial identity to $10^{-4}$ (table~\ref{tab:bounds}), a check that the reference calculations do not report. Battye and
Sutcliffe~\cite{battye2002}
quote an uncertainty of $\pm0.001$ on $E/B$, that is $\pm0.002$ to
$\pm0.004$ on the energies at $B=2$--$4$, and our values lie inside that
band.

\begin{table}[t]
\centering
\caption{Energies in units of $12\pi^2$. The
first row gives the published reference energies $E$, the raw grid energies multiplied by $B$ and divided by the charge computed on the same grid~\cite{battye2002,battye2001}. The second row
gives the energies $E$ of the headline solutions,
evaluated on the $80\times44\times60$ quadrature grid (section~\ref{sec:eval}).
At each charge, the headline solution is the successful run of section~\ref{sec:free} with the lowest energy on this grid. Section~\ref{sec:uniform} gives the network of each headline solution.
The third and fourth
rows give the energies of the first two rows divided by $B$. The last two
rows give the verification metrics of the solutions in the second row.}
\begin{tabular}{lcccc}
\toprule
 & $B=1$ & $B=2$ & $B=3$ & $B=4$ \\
\midrule
Reference, $E$~\cite{battye2002,battye2001} & 1.2322 & 2.3582 & 3.4386 & 4.4804 \\
This work, $E$ & 1.23145 & 2.35697 & 3.43706 & 4.47992 \\
Reference, $E/B$ & 1.2322 & 1.1791 & 1.1462 & 1.1201 \\
This work, $E/B$ & 1.23145 & 1.17848 & 1.14569 & 1.11998 \\
Verification: $|B-n|$ & $<10^{-6}$ & $<10^{-6}$ & $10^{-6}$ & $10^{-6}$ \\
Verification: $E_2/E_4$ & 1.0000 & 0.9999 & 1.0001 & 1.0001 \\
\bottomrule
\end{tabular}
\label{tab:bounds}
\end{table}

(v)~The two architectures differ in how their energies depend on the
evaluation grid. Re-evaluating every successful run of
tables~\ref{tab:uniform} and~\ref{tab:wallclock} ($237$
runs\footnote{The $237$ runs are the $183$ successful runs
of table~\ref{tab:uniform} and the $54$ successful runs of
table~\ref{tab:wallclock} with a fixed time: $14$ of KAN$[16]{\times}4$ at
$160\,$s and $40$ of GELU-MLP$[64]{\times}4$ at $660\,$s.}) on the
$80\times44\times60$ grid, instead of the
$48\times28\times36$ grid used for table~\ref{tab:uniform}, separates the
two architectures (appendix~\ref{app:stats} summarises the re-evaluated energies by network and charge). The MLP energies move by at most $5\times10^{-6}$, with a
median of $3\times10^{-8}$. The KAN energies move by up to
$2.9\times10^{-4}$, in either direction, and the shifts average to zero
across the runs. The two architectures differ in the smoothness of the
fields they produce. A GELU-MLP field is smooth to all
orders, so Gauss--Legendre quadrature converges rapidly on it. By contrast, the KAN field
is built from cubic B-splines, which are twice continuously differentiable
and no more. The energy density is quadratic and quartic in the first
derivatives of the field, so as a function of position it is only once
continuously differentiable: its second derivative jumps across the surfaces
where a spline knot is crossed. Gauss--Legendre quadrature converges slowly
on such an integrand, and the error that remains depends on where the
quadrature nodes fall relative to those surfaces. A change of grid moves
the nodes relative to those surfaces, so the error changes with the grid
and carries no systematic sign. The shift stays below $10^{-4}$ in $96\%$
of the KAN
runs. The KAN energies quoted in this section therefore carry a resolution
uncertainty of $\pm3\times10^{-4}$. The four solutions of
table~\ref{tab:bounds} were obtained with GELU-MLPs and are not affected.

\subsection{Failure modes}
\label{sec:failmodes}
Every run carries its full verification record, so the unsuccessful runs
can be classified rather than discarded (table~\ref{tab:failures}). Three
modes occur. In mode~(a), sector escape, the charge moves away from the
target during relaxation; mode~(a) is the
imprinting-related mode analysed in section~\ref{sec:freeze}. In mode~(b),
unrelaxed configurations, the charge stays at the target charge but the
virial identity~\eqref{eq:virial} fails: $E_2/E_4$ is off by $8$ to
$39\%$. The identity holds at every static solution, so a mode-(b)
configuration is not a stationary point of the energy. In each of the
nine runs in this mode the energy was still falling when the epoch budget
ended. In mode~(c),
high energy, the run passes the verification protocol of
section~\ref{sec:eval} but ends more than $1\%$ above the reference
energy~\cite{battye2002,battye2001}. The mode-(c) configurations are not true local minima, since in each of
the $21$ runs the energy was still falling when the epoch budget ended.
Appendix~\ref{app:modes} gives the rule
that assigns each unsuccessful run to one of the three modes, the counts
for each network, and the properties of the runs in each mode.

\begin{figure}[t]
\centering
\includegraphics[width=\textwidth]{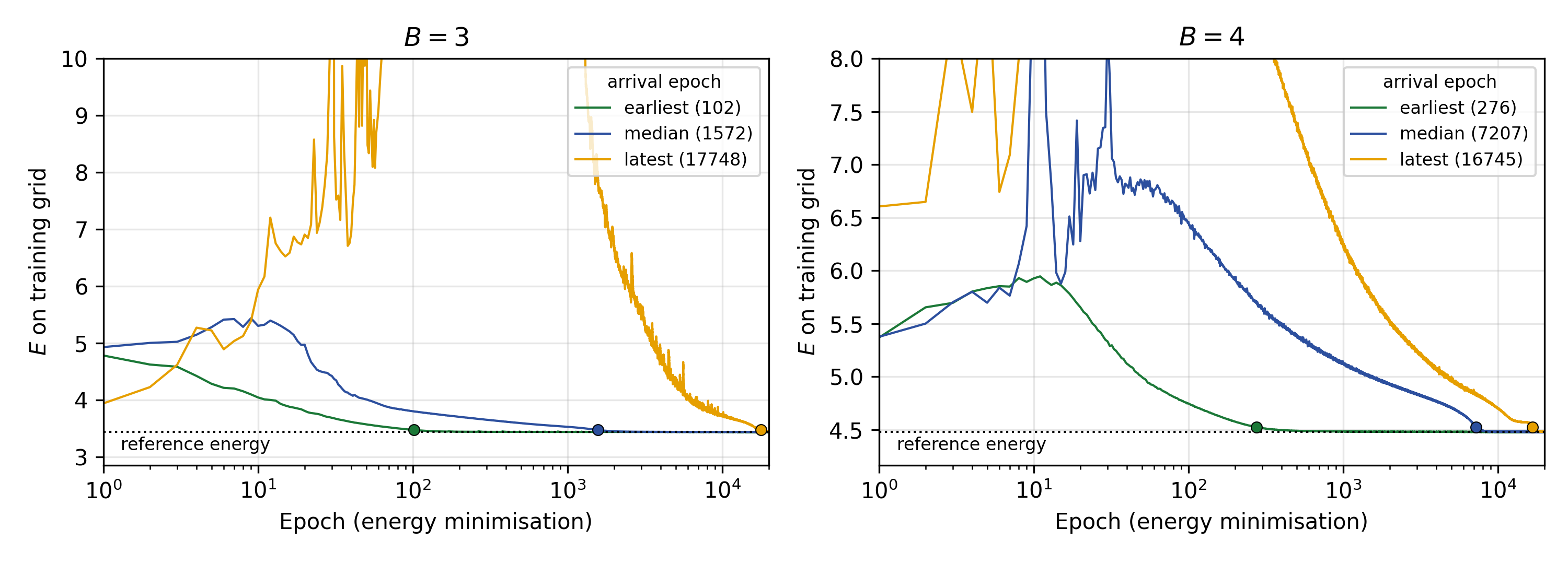}
\caption{Energy on the training grid versus epoch for three successful
uniform-protocol runs at $B=3$ (left) and $B=4$ (right): among the
successful runs of the four networks of table~\ref{tab:uniform}, those
with the earliest, the median and the latest arrival epoch
(section~\ref{sec:failmodes}). The legend gives
the arrival epoch of each run, and a dot marks it on the line. The dotted
line is the published reference energy of table~\ref{tab:bounds}~\cite{battye2002,battye2001}. The
energy fluctuates from one epoch to the next because the nodes of the
training grid move at every training step.}
\label{fig:curves}
\end{figure}

Modes~(b) and~(c) raise the question of how tight the $20\,000$-epoch budget is.
When
GELU-MLP$[64]{\times}4$ is re-trained from the same initial data for about
$80\,000$ epochs, four times the uniform-protocol budget, every failed run
of that network reaches the reference energy. The successful runs give a
second measure of how tight the budget is: how late the slowest of them
reach the reference energy. To measure this, we define the arrival epoch
of a successful run as the first epoch after which the energy on the
training grid stays within $1\%$ of the reference energy. Over the
successful runs of the four networks of table~\ref{tab:uniform}, the
arrival epoch lies between $102$ and
$17\,748$ at $B=3$ ($65$ runs), with a median of $1\,572$. At $B=4$ it
lies between $276$ and $16\,745$ ($39$ runs), with a median of $7\,207$.
Figure~\ref{fig:curves} shows the energy along training for three of
these runs at each charge: the runs with the earliest, the median and the
latest arrival epoch. To summarize, the $20\,000$-epoch budget is tight:
the slowest successful runs need almost all of it to reach the reference
energy.

\begin{table}[t]
\centering
\caption{Classification of the uniform-protocol runs at $B=1$ to $4$ into
successes and the three failure modes. The three failure modes are
defined in the text; appendix~\ref{app:modes} gives the rule that assigns
each unsuccessful run to one of them. Counts are summed over the four
networks of table~\ref{tab:uniform}: $80$ runs per charge at $B\ge3$ and
$40$ at $B\le2$.}
\begin{tabular}{lccccl}
\toprule
 & $B=1$ & $B=2$ & $B=3$ & $B=4$ & diagnostic signature \\
\midrule
(a) sector escape & 0 & 0 & 6 & 21 & $|B-B_{\rm t}|\ge10^{-3}$ \\
(b) unrelaxed & 0 & 0 & 2 & 7 & $|B-B_{\rm t}|<10^{-3}$, $|E_2/E_4-1|=0.08$--$0.4$ \\
(c) high energy & 0 & 1 & 7 & 13 & $|B-B_{\rm t}|<10^{-3}$, protocol passed, $E$ $+1$--$10\%$ \\
\midrule
successful & 40 & 39 & 65 & 39 & all checks passed \\
\bottomrule
\end{tabular}
\label{tab:failures}
\end{table}

\subsection{Equal wall-clock budget}
\label{sec:wallclock}
Even with our implementation, which is $3$ to $24$ times faster than
\texttt{pykan} (appendix~\ref{app:implcost}), a KAN epoch takes about four
times as long as an MLP epoch: $20\,000$ epochs take about $660\,$s for KAN$[16]{\times}4$
and about $160\,$s for the two GELU-MLPs of table~\ref{tab:uniform}. We therefore equalised the wall-clock
time of KAN$[16]{\times}4$ and GELU-MLP$[64]{\times}4$ at $B=3$ and $B=4$,
leaving everything else, including the
imprinting stage, unchanged. Table~\ref{tab:wallclock} reports the result. Capping the
energy-minimisation phase of KAN$[16]{\times}4$ at $160\,$s, the time
GELU-MLP$[64]{\times}4$ needs for $20\,000$ epochs, allows it only about $4\,850$ epochs. The capped KAN$[16]{\times}4$ then
succeeds in only $3$ of its $20$ runs at $B=4$, against $10$ of $20$ for
GELU-MLP$[64]{\times}4$ at the same wall-clock time ($p=0.041$); at $B=3$
it also falls behind, with $11$ against $16$ of $20$ ($p=0.18$). Conversely,
when GELU-MLP$[64]{\times}4$ is given $660\,$s, the time KAN$[16]{\times}4$
needs for $20\,000$ epochs, it succeeds
in all $20$ runs at $B=3$ and in all $20$ at $B=4$. At the same
$660\,$s, GELU-MLP$[64]{\times}4$ succeeds as often as KAN$[16]{\times}4$
with $20\,000$ epochs at $B=3$ ($20$ and $19$ of $20$ runs, $p=1.0$) and
more often at $B=4$ ($20$ and $11$ of $20$ runs, $p=0.0012$). The success-rate advantage of
table~\ref{tab:uniform} is therefore a per-epoch property, and it
disappears when the budget is counted in seconds with present
implementations. Table~\ref{tab:fair} shows the same change for the energies of the
anchored track at $B=2$: the KAN is lowest when the two models run for the
same number of epochs, and the GELU-MLP is lowest when they run for
$30\,$s.

\begin{table}[t]
\centering
\caption{Equal wall-clock comparison in the free track at $B=3$ and $B=4$.
Each entry is the number of successful runs out of the $20$ runs in that
cell. The two middle columns repeat the $20\,000$-epoch runs of
KAN$[16]{\times}4$ and GELU-MLP$[64]{\times}4$ from table~\ref{tab:uniform};
under each network the table gives the number of epochs and the wall-clock
time of the energy-minimisation phase. In every column we fixed one of the
two numbers and measured the other, and the sign $\approx$ marks the
measured one. The two outer columns then allot each network the wall-clock
time that the other network needed for its $20\,000$ epochs: the first
column runs KAN$[16]{\times}4$ for the $160\,$s of the third column, and the
last column runs GELU-MLP$[64]{\times}4$ for the $660\,$s of the second
column.
Two-sided Fisher exact tests are in the text.}
\begin{tabular}{ccccc}
\toprule
 & KAN$[16]{\times}4$ & KAN$[16]{\times}4$ & GELU-MLP$[64]{\times}4$ &
GELU-MLP$[64]{\times}4$ \\
 & $\approx4\,850$ ep & $20\,000$ ep & $20\,000$ ep & $\approx79\,500$ ep \\
$B$ & $160\,$s & $\approx660\,$s & $\approx160\,$s & $660\,$s \\
\midrule
3 & 11/20 & 19/20 & 16/20 & 20/20 \\
4 & 3/20 & 11/20 & 10/20 & 20/20 \\
\bottomrule
\end{tabular}
\label{tab:wallclock}
\end{table}

\section{Results III: topological charge under gradient descent}
\label{sec:freeze}

\begin{figure}[t]
\centering
\includegraphics[width=0.78\textwidth]{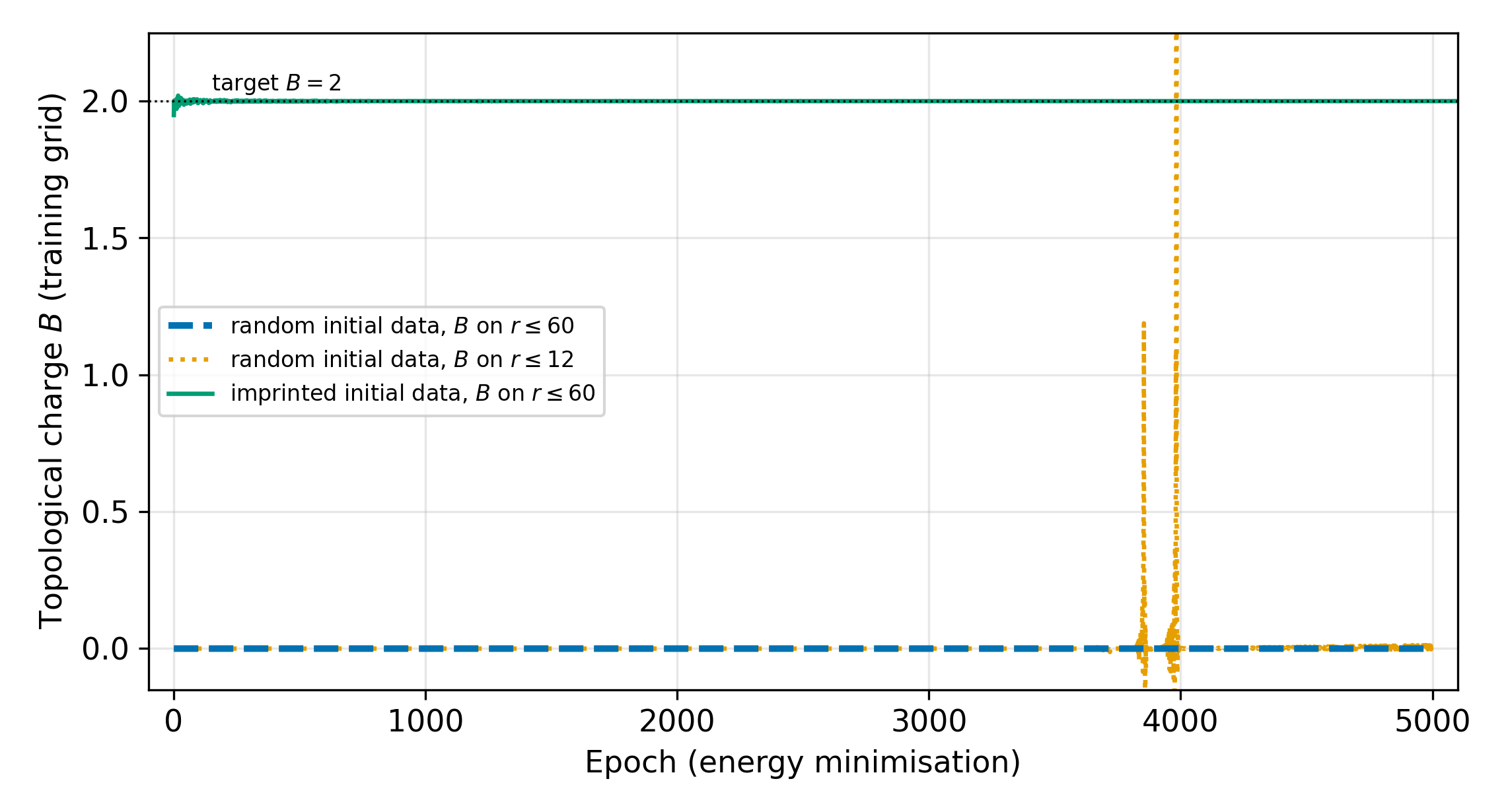}
\caption{Charge trajectories during training (free track, target charge
$B_{\rm t}=2$, KAN $[16]{\times}4$). The vertical axis is the charge $B$
measured on the training grid. The blue dashed line and the yellow dotted
line are $5000$-epoch runs that start from random initial data and hold the
far field at the vacuum through the far-field penalty. The blue dashed run
measures $B$ on $r\le60$, and the yellow dotted run measures $B$ on
$r\le12$. The green solid line is a $10\,000$-epoch run in which
the imprinting step of section~\ref{sec:imprint} fits the network to the
multi-lump initial data before the relaxation. The horizontal axis counts
the epochs of the energy minimisation, so the imprinting of the green run
is not shown, and it ends at epoch $5000$, where the two random-start
runs end; in the rest of the green run $B$ stays within $0.001$ of $2$.
The vertical axis is cut at
$2.25$: the yellow dotted line leaves the frame at epoch $3983$, where $B$
reaches $17.2$. The text explains the three runs.}
\label{fig:freeze}
\end{figure}

\subsection{The mechanism}
\label{sec:freezemech}
We now explain why the imprinting step is necessary.
Section~\ref{sec:methods} states that $B$ is the degree of a
map $S^3\to S^3$ once the far field is at the vacuum, and that a degree does not
change under a continuous deformation of the field. A related observation
has been made in neural network field theory: a network whose output is a
smooth, single-valued field carries no winding number, so the
winding sectors are added as discrete parameters~\cite{ferko2026}. To our
knowledge, however, no earlier work has used the degree argument to explain why a
neural variational solver cannot move between topological sectors except where the field becomes singular. In the
representation $q=\mathbf{v}/\|\mathbf{v}\|$ the charge can change only
where the raw output $\mathbf{v}$ passes through zero, which is a singular
point of the normalisation. Changes of the topological sector are therefore
localised, discrete events rather than gradual drift.
Appendix~\ref{app:modes} reports two re-runs in which we recorded how close
$\mathbf{v}$ came to zero when the charge changed.

\subsection{Extra runs}
The runs of this section lie outside the uniform protocol.
Subsection~\ref{sec:freezeres} uses sixteen runs to ask whether the
charge penalty alone can supply the target charge, and
subsection~\ref{sec:imprintres} uses a designed experiment of fifty
imprinted runs to find how the architecture acts on an imprinted run.
Each subsection describes its runs before its results.

\subsection{Freezing}
\label{sec:freezeres}
Subsection~\ref{sec:freezemech} argues that a training step cannot change
the charge except where the field becomes singular. This subsection tests
whether the charge penalty alone can nevertheless pull a run into the
target sector.

\paragraph{Runs.}
We use sixteen runs in which the target charge enters only through the
charge penalty of equation~\eqref{eq:loss}. They form the four groups of
table~\ref{tab:freezeruns}. Groups 1 and 2 use the anchored construction with
the background $q_{\rm base}$, which gives the target charge in the anchored
track, replaced by the constant $q_{\rm vac}$, which carries charge zero; the
far field is still at the vacuum by construction.\footnote{Unlike the anchored track of section~\ref{sec:field},
these runs initialise the last layer of the network at random rather than to
zero. With a zero last layer the field would be exactly $q_{\rm vac}$, a
stationary point of the loss, and training would not move.} Groups 3 and 4
use the free track without the imprinting step, which gives the target
charge in a free run (section~\ref{sec:imprint}).

The four groups differ in the conditions under which the charge penalty
acts.
\begin{description}
\item[Group 1] is the plain case: the anchored-track weight $w_{\rm topo}=200$
and no further term.
\item[Group 2] raises $w_{\rm topo}$ to the free-track value
$1000$ and adds a term against collapse to the vacuum,
$\tfrac{1}{4}\,w_{\rm origin}\langle\|q+q_{\rm vac}\|^2\rangle$ with $w_{\rm origin}=10$,
evaluated on $64$ random points near the origin: $-q_{\rm vac}$ is the value
that the rational-map ansatz~\eqref{eq:rmap} takes at the origin for every
$B$, because $f(0)=\pi$, so the term holds the field near the origin away
from the vacuum.
\item[Group 3] keeps $w_{\rm topo}=1000$ and $w_{\rm origin}=10$, replaces the construction
$\mathcal{N}[q_{\rm vac}+g_{\rm soft}\mathbf{v}]$ by the free-track field
$\mathcal{N}[\mathbf{v}]$, and adds the far-field term of
section~\ref{sec:loss} with $w_{\rm far}=50$ to hold the far field at the
vacuum. Group 3 tests whether the outcome depends on the anchored construction.
\item[Group 4] differs from group 3 only in where $B$ is measured for the charge penalty
$w_{\rm topo}(B-B_{\rm t})^2$: on the ball $r\le12$ instead of $r\le60$. The
degree argument needs $q\approx q_{\rm vac}$ on the boundary of the region
where $B$ is measured (section~\ref{sec:methods}). The far-field term
acts only on points with $30\le r\le60$, not on the sphere $r=12$, so $B$
on the ball $r\le12$ is not quantised. Group 4 thus tests whether $B$
moves when it is not held at an integer.
\end{description}

\begin{table}[h]
\centering
\caption{The sixteen runs of subsection~\ref{sec:freezeres}. The columns
give the construction of the field, how the far field is held at the
vacuum, the weights $w_{\rm topo}$ of the charge penalty and $w_{\rm origin}$
of the origin term, and the ball on which $B$ is measured, as described
group by group in the text. Each
group is four runs: a KAN $[16]{\times}4$ and a tanh-MLP $[64]{\times}4$, at
$B_{\rm t}=1$ and at $B_{\rm t}=2$. None of the sixteen uses imprinting. All sixteen
ran for $5000$ epochs, with Adam, the cosine
schedule and the learning rate $2\times10^{-3}$ of section~\ref{sec:protocol}.
Groups 1 and 2
replace the background $q_{\rm base}$ of the anchored construction by the
constant $q_{\rm vac}$, with the envelope $g_{\rm soft}(r)=3(1-s)^2$ and
$s=r/(3+r)$, which decays as $r^{-2}$ at infinity.}
\begin{tabular}{cllccc}
\toprule
Group & Construction & Far field & $w_{\rm topo}$ & $w_{\rm origin}$ & $B$ measured on \\
\midrule
$1$ & $q_{\rm vac}+g_{\rm soft}\mathbf{v}$ & by construction & $200$ & $0$ & $r\le60$ \\
$2$ & $q_{\rm vac}+g_{\rm soft}\mathbf{v}$ & by construction & $1000$ & $10$ & $r\le60$ \\
$3$ & free track, $\mathbf{v}$ alone & penalty, $w_{\rm far}=50$ & $1000$ & $10$ & $r\le60$ \\
$4$ & free track, $\mathbf{v}$ alone & penalty, $w_{\rm far}=50$ & $1000$ & $10$ & $r\le12$ \\
\bottomrule
\end{tabular}
\label{tab:freezeruns}
\end{table}

\paragraph{Results.}
All twelve runs of groups 1 to 3 ended with $|B|$ below $10^{-4}$.
None of the twelve reached its target sector.
Six of the twelve reached the vacuum,
and the other six ended at an energy between $0.15$ and $0.60$ at
$B_{\rm t}=1$ and between $0.15$ and $1.30$ at $B_{\rm t}=2$, against the
reference values $1.232$ at $B=1$ and $2.358$ at $B=2$~\cite{battye2002,battye2001}. The blue dashed line of
figure~\ref{fig:freeze} is the KAN of group $3$ at $B_{\rm t}=2$, and $|B|$ stays below $10^{-6}$ for all $5000$ epochs.
The charge did move in one of the four runs of group 4, the KAN at
$B_{\rm t}=2$, which is the yellow dotted line. There $|B|$ stayed below
$10^{-3}$ until epoch $3665$, rose in two bursts to $1.19$ and then to
$17.19$, and from epoch $3988$ stayed below $0.05$ for the rest of the run,
ending at $B=0.010$. In the other three runs of group 4, $|B|$ stayed below
$3\times10^{-4}$ after the first twenty epochs. None of the four reached its
target sector. The green solid line of figure~\ref{fig:freeze}, by contrast,
is an imprinted run, and $B$ stays within $0.06$ of $2$ for all $10\,000$
epochs. We use the
word freezing after lattice Monte Carlo, where the topological charge of
gauge configurations changes more and more rarely as the continuum limit
is approached~\cite{deldebbio2002,schaefer2011,luscher2010,kanwar2020},
whereas here the degree changes only where the field becomes singular.

\subsection{Imprinting}
\label{sec:imprintres}
The charge penalty alone pulled none of the sixteen runs of
subsection~\ref{sec:freezeres} into its target sector. This suggests that
the field should already carry the target charge when the relaxation
starts. That is what the imprinting step of section~\ref{sec:imprint} is
for. This subsection asks how the architecture acts on an imprinted run.

\paragraph{Runs.}
We use a designed experiment of fifty runs, all at $B_{\rm t}=2$. Forty-five of the fifty vary how far the imprinting
regression of section~\ref{sec:imprint} is taken. Instead of running for the
fixed $8000$ steps of the uniform protocol, the regression is stopped as soon
as the imprinting loss, averaged over the last $50$ steps, falls below a
target value. We use three target
values, $10^{-3}$, $10^{-4}$ and $10^{-5}$. Three networks are run
with each target at five seeds: a KAN $[16]{\times}4$, a GELU-MLP $[64]{\times}4$
and a tanh-MLP $[64]{\times}4$. The other five runs are
tanh-MLPs whose regression stops after those $8000$ steps, instead of at a
target imprinting loss. We add them to
find whether the tanh-MLP fails under the uniform protocol because its
imprinting regression is too short or because of the network itself. In all fifty runs the relaxation that follows the
imprinting uses the uniform protocol. After the regression, we evaluate the
imprinting loss once more, averaged over a fixed set of points that is the
same for all fifty runs.

\paragraph{Results.}
What decides the outcome is how far the imprinting regression is taken, and
not which network runs it. At a target imprinting loss of $10^{-3}$ no run
of any of the three networks succeeded. At $10^{-4}$ each network succeeded
in two of five runs, though not on the same seeds, and at $10^{-5}$ in five
of five (table~\ref{tab:imprintdesign}). Hence the three networks succeeded in the
same number of runs at every target imprinting loss (Fisher exact $p=1.00$
for every pair at a common target imprinting loss).
Summing these counts over the three networks, the fifteen runs at each
target imprinting loss gave $0$, $6$ and $15$ successes as the target
imprinting loss falls
(Fisher exact $p=0.017$ between $10^{-3}$ and $10^{-4}$, and
$p=7\times10^{-4}$ between $10^{-4}$ and $10^{-5}$). Four of the
forty-five runs in table~\ref{tab:imprintdesign} met the charge criterion
but ended at an energy between $2.50$ and $3.30$, above the reference value
$2.358$ at $B=2$~\cite{battye2002,battye2001}. Two of the four are mode~(b) of
table~\ref{tab:failures} and the other two are mode~(c).

The networks differ in what a target imprinting loss costs. Reaching a
target imprinting loss of $10^{-5}$ took the KAN
$2587$ to $3724$ regression steps and the GELU-MLP $8283$ to $10\,479$,
while the tanh-MLP needed more than $29\,000$; three of the five tanh runs
stopped at the cap of $30\,000$ steps, at an imprinting loss between
$2.4\times10^{-5}$ and $3.4\times10^{-5}$. The uniform protocol gives the
regression $8000$ steps, far fewer than the tanh-MLP needs for $10^{-5}$.
The other five runs of the fifty, the tanh-MLPs that stopped at those
$8000$ steps, reached an imprinting loss between $6.9\times10^{-4}$ and
$8.5\times10^{-4}$ only, and none of the five succeeded. The difference
between the networks is therefore the
cost of the fit, and not what the network can reach: once the regression of
the tanh-MLP is not cut off at the $8000$ steps of the uniform protocol,
it succeeds as often as the other two.

\begin{table}[h]
\centering
\caption{Outcome of the forty-five runs that stop the imprinting regression
at a target imprinting loss (subsection~\ref{sec:imprintres}). The other
five runs, which stop at the $8000$ steps
of the uniform protocol, are described in the text. Each entry is a number of runs
out of the five seeds of that cell, and all runs are at $B_{\rm t}=2$.
The success columns count successful runs in the sense of
section~\ref{sec:eval}, as in table~\ref{tab:uniform}, and the charge
columns count runs that ended with $|B-B_{\rm t}|<10^{-3}$, whether or not
they also succeeded. The
imprinting regression stops when the mean imprinting loss over the last
$50$ regression steps first falls below the target imprinting loss. In
addition the regression stops at $30\,000$ steps if the mean does not fall
below it. Everything after the imprinting follows the uniform protocol.}
\begin{tabular}{lcccccc}
\toprule
 & \multicolumn{2}{c}{KAN $[16]{\times}4$}
 & \multicolumn{2}{c}{GELU-MLP $[64]{\times}4$}
 & \multicolumn{2}{c}{tanh-MLP $[64]{\times}4$} \\
\cmidrule(lr){2-3}\cmidrule(lr){4-5}\cmidrule(lr){6-7}
Target imprinting loss & charge & success & charge & success & charge &
success \\
\midrule
$10^{-3}$ & $0$ & $0$ & $0$ & $0$ & $1$ & $0$ \\
$10^{-4}$ & $3$ & $2$ & $3$ & $2$ & $3$ & $2$ \\
$10^{-5}$ & $5$ & $5$ & $5$ & $5$ & $5$ & $5$ \\
\bottomrule
\end{tabular}
\label{tab:imprintdesign}
\end{table}

\section{Discussion}
\label{sec:discussion}

\subsection{Implications for the KAN--MLP debate}
\label{sec:kanmlp}
Our benchmark reproduces both findings reported in the literature: KANs
outperform MLPs in some studies and underperform them in others. Against a tanh-MLP at default settings the KAN
reaches an energy lower by $9\times10^{-4}$ (table~\ref{tab:fair}). Against
a tuned GELU-MLP, that is a GELU-MLP at its best learning rate, the
difference is smaller than $10^{-4}$. Whether the KAN or the
GELU-MLP is ahead depends on whether the budget is counted in epochs or
in seconds (table~\ref{tab:fair}). A published gap between a KAN and an MLP can therefore have two sources
at once: the architectures themselves, and the settings with which they
are trained, such as the activation function and the learning rate
(section~\ref{sec:limitations}). The result of an MLP depends on the activation
function of the MLP, which is often left at the default value.

For practitioners the summary is as follows. On this benchmark, KANs are the safer default. KANs
are more parameter-efficient between about $4\times10^2$ and $10^4$
parameters (figure~\ref{fig:pareto}(a)), and KANs tend to succeed more often
at equal epochs. KANs do not succeed more often at equal wall-clock time on
present-day hardware (table~\ref{tab:wallclock}). The wall-clock time
itself depends on the implementation. Our own code is $3$ to $24$ times faster
than \texttt{pykan} (appendix~\ref{app:implcost}), and a JAX
re-implementation of physics-informed KANs is reported to run $67$ to $85$
times faster than \texttt{pykan} per training epoch~\cite{rigas2024}. Reported speed gaps between KANs
and MLPs therefore depend at least as much on the code that runs the
network as on the architecture.

For benchmark designers, a field with a topological charge gives a quantity
whose correct value is known before the run, even if the solution itself
is not known in closed form. A measured charge that differs from this
value therefore signals a wrong result. In our runs this check caught two
kinds of wrong result: runs that left the
target sector, and runs of the \texttt{FastKAN} variant of
section~\ref{sec:arch} in which the quadrature under-counts the energy
(appendices~\ref{app:modes} and~\ref{app:quadfail}).

\subsection{Physics outlook}
The ansatz-free solver needed no symmetry input for $B=2$ to $4$. The
known toroidal, tetrahedral and cubic symmetries emerged on their own
(section~\ref{sec:symmetry}). The solver also produced variational upper
bounds for $B=1$ to $4$, which were checked on evaluation grids that the
training never used. The relative orientations of the lumps in the initial
data were also left random (section~\ref{sec:imprint}). Choosing
orientations in which the lumps attract one another could shorten the
relaxation.

There are three natural next steps: $B\ge5$, the massive-pion
model~\cite{battye2005}, and generalised Skyrme models with sextic or
BPS-type terms~\cite{adam2010}. At $B\ge5$ the number of local minima grows
and their energies lie close together~\cite{feist2013,gudnason2022}. Because
our method assumes no symmetry, the same method applies to all three.
A comparison with existing solvers is a separate question. We
did not re-implement a grid-based or gradient-flow solver, so a comparison
at matched cost is left as future work.

Skyrmion scheme is one of the fundamental bases for constructing nuclear physics; once quantized, Skyrmion bound states represent nuclear energy levels. In addition, classical local minima of the Skyrmion solution landscape should correspond to long-lived excited states of atomic nuclei. Therefore, our neural method based on no symmetry ansatz with secured baryon number sectors can be an opening door to the entire Skyrmionic nuclear physics. We believe that the topological benchmark provides a ground not only for a healthy competition for precision physics but also for a future connection between hadron physics and nuclear physics.

\subsection{Limitations}
\label{sec:limitations}
Four limitations bound our claims. (i)~Benchmark scope. Every conclusion
comes from the Skyrme model alone. Section~\ref{sec:kanmlp} concludes that
a published gap between a KAN and an MLP can have two sources at once:
the architectures, and the settings with which they are trained. We
expect the conclusion to hold for other localised variational problems, but
we did not test other problems here.
KANs were proposed in part because they are interpretable: the function
on each edge can be plotted and replaced by a formula~\cite{liu2024kan}.
For a physics problem whose exact solution has a closed form,
interpretability could give KANs an advantage over MLPs, because a KAN
could then represent the solution exactly with formulas on its
edges~\cite{liu2024kan}. No Skyrmion of the Skyrme model of
section~\ref{sec:model} is known in closed form~\cite{manton2004}, so our
benchmark cannot test whether interpretability gives KANs an advantage
over MLPs.

(ii)~Statistics. In the uniform-protocol study of table~\ref{tab:uniform},
each network was run from twenty random placements at $B=3$ and $B=4$,
against ten at $B=1$ and $B=2$. The decision to add the second ten came
after the result of the first ten had been seen. The second ten were added
for statistical power, and every run is reported. Adding runs after a
result has been seen is a form of optional stopping, which raises the
chance of calling a difference significant when no difference is present.
The $p$-values of section~\ref{sec:free} may therefore be smaller than
they would be if the twenty runs had been planned from the start. There is another issue, caused by the number
of $p$-values that we report. Sections~\ref{sec:free}
and~\ref{sec:freeze} report thirteen $p$-values in all, so a single value
near $0.05$ is weak evidence on its own.
A Bonferroni correction over the thirteen $p$-values
sets the threshold at $0.05/13=0.0038$. The effect of the target imprinting
loss (table~\ref{tab:imprintdesign}) and the effect of the wall-clock
budget (table~\ref{tab:wallclock}) have $p=7\times10^{-4}$ and $p=0.0012$,
respectively, and are below the threshold. The difference in success rate between KAN$[16]{\times}4$ and
GELU-MLP$[48]{\times}4$ at $B=3$ has $p=0.044$ and is above the threshold,
so we present it as a trend rather than as a definitive effect.

(iii)~Training length. Within the
$20\,000$-epoch protocol each of the four networks succeeded in six to
twelve of its twenty runs at $B=4$ (table~\ref{tab:uniform}).
As we saw in section~\ref{sec:wallclock}, when given about four times
more epochs, GELU-MLP$[64]{\times}4$ succeeded in all twenty of its runs
at $B=4$ (table~\ref{tab:wallclock}). Hence the success rates of
table~\ref{tab:uniform} depend on the training length.

(iv)~Settings that we did not tune. Both architectures have
many more settings than those that we varied, such as the width, the depth, the
learning rate and the activation of the MLP. Some of the settings that we
did not vary belong to both architectures, such as the encoding of the
input coordinates~\cite{tancik2020}.
Others belong to one architecture only, such as the number $G$ of grid intervals, the spline
degree $k$ and the basis functions of the KAN. We kept these settings at
the values of section~\ref{sec:arch} and appendix~\ref{app:inventory},
which includes the initialisation of the weights and the learning-rate
schedule.
Each of them can be tuned, and tuning a setting of one architecture can
move the comparison towards that architecture. For example, changing the
activation of the MLP moved it towards the MLP (section~\ref{sec:fair}).
On the KAN side, we also
made $12$ runs at $G=3$ and $G=10$ to see how its energy depends on $G$,
and the energy fell as $G$ grew. A comparison of the two architectures
therefore holds only for the settings that were tuned.

\subsection{Reproducibility}
\label{sec:repro}
Every run in this study, including every failure, is stored as a
self-contained record. The record holds the full hyperparameter settings,
the random seed, the per-epoch histories of energy and
charge, and all final verification metrics. Tables and figures are generated
directly from the records by scripts, so each
number in the paper traces back to on-disk evidence. Residual GPU
nondeterminism was measured rather than assumed. We re-ran thirteen runs
with their original seeds: three free-track runs of the uniform protocol
(KAN$[16]{\times}4$
at $B=2$ and $B=4$, GELU-MLP$[64]{\times}4$ at $B=3$) and ten runs of the
anchored sweep at $B=2$.\footnote{The ten re-runs repeat five runs of
the anchored sweep at $B=2$: KAN$[2]{\times}4$, KAN$[16]{\times}1$ and
KAN$[64]{\times}3$ at seed~$3$, and tanh-MLP$[8]{\times}4$ and
tanh-MLP$[32]{\times}2$ at seed~$4$. We ran each of the five runs twice
more with the same seed and the same code, and compared each re-run with
the run of the anchored sweep that has the same settings and
seed.} Nine of the thirteen were reproduced bit for bit, with identical
energy, charge and virial ratio. Of the other four, the free-track run
KAN$[16]{\times}4$ at $B=2$ differed by $1.6\times10^{-5}$ in $E$. A
difference of $1.6\times10^{-5}$ is smaller than every energy difference
that table~\ref{tab:fair} marks in bold. The other three, all anchored
runs, differed by $2.1\times10^{-8}$ (the two re-runs of KAN$[16]{\times}1$)
and $1.5\times10^{-4}$ (a width-$8$ tanh-MLP). Networks of width $8$
are not converged at the $1000$ epochs of the anchored track of
section~\ref{sec:protocol}, which makes them sensitive to small
numerical differences. Moreover, the difference of $1.5\times10^{-4}$ is smaller
than the change that a new seed makes: in figure~\ref{fig:pareto} the seed
standard deviations of the tanh-MLP have a median of $7.7\times10^{-4}$.

\subsection{Honest reporting}
Success
rates are always reported for the complete cohort of a
protocol, with no post-hoc selection of runs. The uniform protocol of
table~\ref{tab:uniform} was fixed before any of its runs were launched.
Every run of the experiments in this paper is reported, $1\,398$ runs in
all, including those that failed.

\section{Conclusion}
\label{sec:conclusion}
We have introduced the Skyrme model as a topologically nontrivial
benchmark for neural field methods, in which a result can be checked
against exact relations. To our knowledge, we have obtained the first neural solutions for
$B=1$--$4$ that are fully three-dimensional and use no symmetry ansatz. At $B=2$--$4$ the known symmetries emerge on
their own. The variational energies lie slightly below, and are
consistent with, the published reference values~\cite{battye2002,battye2001}. Using the benchmark we have
shown that most of the performance gap between KANs and MLPs on this
benchmark disappears once the activation function and learning rate of the MLP are
tuned. Three KAN advantages remain. (i)~The KAN is more
parameter-efficient than the GELU-MLP between about $4\times10^2$ and
$10^4$ parameters. (ii)~At a matched count of about $8\times10^3$ parameters
and equal epoch budget the KAN tends to have a higher success rate than
the GELU-MLP. (iii)~The KAN is more robust to default settings than the MLP.

We applied the verification protocol to every run reported in this paper.
The protocol has four parts: charge quantisation, the virial identity,
the Faddeev--Bogomolny bound $E\ge|B|$, and the spread of the energy over
independent quadrature grids of the same resolution. We
recommend the protocol as standard practice for physics-informed learning on
topological problems.

\section*{Acknowledgments}
The authors designed the study and ran the $1\,398$ training runs that
this paper reports. Every number in the paper was checked against the run
records. In writing the code and the text of this paper, the authors used
Claude Fable 5, Claude Fable 5.1, Claude Opus 5 and Claude Opus 5.5
(Anthropic) as assistants. The physical quantities computed by the code
agree with an independent implementation written by the authors. The
authors take responsibility for all material in this paper.

The work of K.~H.\ was supported in part by JSPS KAKENHI Grant No.\
JP22H05111 and JP22H05115, and by JST, ARiSE, Japan Grant Number JPMJAR2633.
The work of M.~M.\ and G.~O.\ was
supported in part by JSPS KAKENHI Grant No.\ JP22H05115.

\section*{Data availability}
The data that support the findings of this study are openly available at
\url{https://doi.org/10.5281/zenodo.24390286}. The data contain the source
code, the settings, the records and the per-epoch training histories of
every run that this paper reports, including the runs that failed.
The source code and the records can also be read at
\url{https://github.com/MasakiMurata1982/neural-variational-skyrmions}.
The reference energies used for comparison are taken from refs.~\cite{battye2002,battye2001}.

\appendix

\section{Verification and statistics}
\label{app:verif}
The five subsections below give the numbers behind the verification
protocol: the metrics of the four headline solutions, the charge
of the \texttt{FastKAN} runs and of the anchored sweep, the energies reached by each network at each
charge, the classification of the failures, and a quadrature failure of
the \texttt{FastKAN} variant.

\subsection{Headline solutions}
\label{app:headline}
This appendix reports how far the four headline solutions
(section~\ref{sec:uniform}) lie inside the thresholds of the verification
protocol. Table~\ref{tab:headline} gives the solutions and the
verification metrics of each one. Every value lies far inside the threshold of the protocol:
$|B-n|$ is at most $10^{-6}$ against a threshold of $10^{-3}$,
$|E_2/E_4-1|$ at most $10^{-4}$ against $0.05$, and $\Delta E$ at most
$6.4\times10^{-7}$ against $10^{-2}$.
\begin{table}[h]
\centering
\caption{Energy, verification metrics and symmetry of the headline
(lowest-energy) solutions, evaluated on the $80\times44\times60$ grid. The
number in the last column is the relative difference between the principal
moments~\eqref{eq:inertia} that the symmetry requires to be equal:
$(I_3-I_1)/I_1$ for the spherical, tetrahedral and cubic solutions and
$(I_2-I_1)/I_1$ for the torus. The principal moments are evaluated not on
the $80\times44\times60$ grid but on a $48\times28\times36$ grid with
$r_0=3$, truncated at $r=30$.}
\label{tab:headline}
\begin{tabular}{lccccc}
\toprule
Solution & $E$ & $|B-n|$ & $E_2/E_4$ & $\Delta E$ & symmetry (degeneracy) \\
\midrule
$B{=}1$, GELU-MLP$[48]{\times}4$, seed 4 & 1.23145 & $<10^{-6}$ & 1.0000 & $2.0{\times}10^{-7}$ & spherical ($1.1{\times}10^{-4}$) \\
$B{=}2$, GELU-MLP$[64]{\times}4$, seed 7 & 2.35697 & $<10^{-6}$ & 0.9999 & $4.5{\times}10^{-7}$ & toroidal ($8{\times}10^{-5}$) \\
$B{=}3$, GELU-MLP$[64]{\times}4$ $660\,$s, seed 13 & 3.43706 & $10^{-6}$ & 1.0001 & $6.0{\times}10^{-7}$ & tetrahedral ($8{\times}10^{-5}$) \\
$B{=}4$, GELU-MLP$[64]{\times}4$ $660\,$s, seed 16 & 4.47992 & $10^{-6}$ & 1.0001 & $6.4{\times}10^{-7}$ & cubic ($1.3{\times}10^{-4}$) \\
\bottomrule
\end{tabular}

\end{table}

\subsection{Charge of the \texttt{FastKAN} runs and of the anchored sweep}
\label{app:charge}
This appendix reports whether the anchored runs kept the charge of
$q_{\rm base}$. The runs are the $600$ runs of the anchored sweep and the
$144$ anchored runs of \texttt{FastKAN}.
The anchored construction preserves the charge as long as
$\|g\,\mathbf{v}\|<1$ pointwise. Indeed, since $\|q_{\rm base}\|=1$, the
straight-line path from $q_{\rm base}$ to $q_{\rm base}+g\mathbf{v}$
never reaches zero, so $\mathcal{N}$ is defined along it and $q$ is a
continuous deformation of $q_{\rm base}$. The condition $\|g\,\mathbf{v}\|<1$ is not enforced
during training, so we monitor $B$ itself.

All $600$ runs of the anchored sweep (section~\ref{sec:fair}) pass all four
checks of the verification protocol of section~\ref{sec:eval}. Their $|B-n|$ has median
$3\times10^{-7}$ and stays below $1.8\times10^{-5}$.

\texttt{FastKAN} was run in the anchored track at the $24$ (width, depth)
pairs of the anchored sweep and three seeds, at
$B=1$ and $B=2$ ($144$ runs). Of the $144$ runs, $125$ pass two
checks of the verification protocol, $\Delta E<10^{-2}$ and $|E_2/E_4-1|<0.05$. Among the $125$ runs, $|B-n|$ has median
$6.5\times10^{-5}$, and three runs exceed the $10^{-3}$ tolerance, reaching
$1.5\times10^{-3}$. All three have depth $1$ or $2$. Including the $19$
runs that fail one or both of the two checks, the nearest integer to $B$ is
the target charge in all $144$ runs: the largest deviation from it is
$1.1\times10^{-2}$.

For comparison, the free track has no background $q_{\rm base}$, so $B$
is not guaranteed to stay at the target charge.
Failure mode~(a) of table~\ref{tab:failures} collects the uniform-protocol runs
that miss the target charge.

\subsection{Energy statistics by network and charge}
\label{app:stats}
This appendix reports how much the energies of the four networks of
table~\ref{tab:uniform} differ at each charge.
Table~\ref{tab:stats} gives the energies of the successful runs of
table~\ref{tab:uniform} and of the $660\,$s runs of table~\ref{tab:wallclock}
for each network and charge, re-evaluated on the
$80\times44\times60$ grid. Three points follow. (i)~At $B=1$ and $B=2$ the four networks
reach the same energy: their mean energies differ by at most
$1.5\times10^{-5}$, and their lowest energies differ by at most
$1.2\times10^{-5}$. (ii)~At $B\ge3$ the mean energies of the four networks
differ by up to $1.2\times10^{-3}$, while their lowest energies differ by at
most $4\times10^{-5}$ at $B=3$ and $4\times10^{-4}$ at $B=4$. (iii)~The $660\,$s GELU-MLP$[64]{\times}4$ runs reach about
$8\times10^4$ epochs, and their energies have a standard deviation of only
$1.3\times10^{-4}$ at $B=4$, against $1.5\times10^{-3}$ at $20\,000$ epochs.
We therefore expect the runs of each network to approach the same energy
as the number of epochs grows.

\begin{table}[h]
\centering
\caption{Energy of the successful free-track runs of each network at each charge,
in units of $12\pi^2$. The columns give the number of successes out of the
number of runs, and the mean, standard deviation and minimum of the energy
on the $80\times44\times60$ grid. The last column is the median wall-clock
time over all runs of the cell, including the imprinting. The first sixteen
rows are the four networks of table~\ref{tab:uniform}, whose energy
minimisation stops after $20\,000$ epochs. The last two rows are the
GELU-MLP$[64]{\times}4$ runs of table~\ref{tab:wallclock} with $660\,$s,
which exist only at $B=3$ and $B=4$.}
\begin{tabular}{clccccc}
\toprule
$B$ & network & successes & mean $E$ & std & min $E$ & time [s] \\
\midrule
1 & KAN $[16]{\times}4$ & $10/10$ & 1.23146 & $7.0{\times}10^{-6}$ & 1.23145 & 684 \\
1 & KAN $[21]{\times}4$ & $10/10$ & 1.23148 & $1.7{\times}10^{-5}$ & 1.23146 & 716 \\
1 & GELU-MLP $[64]{\times}4$ & $10/10$ & 1.23147 & $1.6{\times}10^{-5}$ & 1.23145 & 175 \\
1 & GELU-MLP $[48]{\times}4$ & $10/10$ & 1.23147 & $1.5{\times}10^{-5}$ & 1.23145 & 172 \\
\midrule
2 & KAN $[16]{\times}4$ & $10/10$ & 2.35702 & $9.4{\times}10^{-5}$ & 2.35697 & 692 \\
2 & KAN $[21]{\times}4$ & $10/10$ & 2.35701 & $2.7{\times}10^{-5}$ & 2.35698 & 718 \\
2 & GELU-MLP $[64]{\times}4$ & $10/10$ & 2.35701 & $7.5{\times}10^{-5}$ & 2.35697 & 176 \\
2 & GELU-MLP $[48]{\times}4$ & $9/10$ & 2.35702 & $1.0{\times}10^{-4}$ & 2.35697 & 174 \\
\midrule
3 & KAN $[16]{\times}4$ & $19/20$ & 3.43847 & $3.1{\times}10^{-3}$ & 3.43710 & 701 \\
3 & KAN $[21]{\times}4$ & $17/20$ & 3.43732 & $4.1{\times}10^{-4}$ & 3.43710 & 730 \\
3 & GELU-MLP $[64]{\times}4$ & $16/20$ & 3.43780 & $1.8{\times}10^{-3}$ & 3.43709 & 179 \\
3 & GELU-MLP $[48]{\times}4$ & $13/20$ & 3.43780 & $1.7{\times}10^{-3}$ & 3.43713 & 176 \\
\midrule
4 & KAN $[16]{\times}4$ & $11/20$ & 4.48084 & $1.1{\times}10^{-3}$ & 4.48003 & 699 \\
4 & KAN $[21]{\times}4$ & $12/20$ & 4.48052 & $3.9{\times}10^{-4}$ & 4.48015 & 732 \\
4 & GELU-MLP $[64]{\times}4$ & $10/20$ & 4.48165 & $1.5{\times}10^{-3}$ & 4.48005 & 180 \\
4 & GELU-MLP $[48]{\times}4$ & $6/20$ & 4.48144 & $1.1{\times}10^{-3}$ & 4.48044 & 178 \\
\midrule
\midrule
3 & GELU-MLP $[64]{\times}4$, $660\,$s & $20/20$ & 3.43707 & $2.4{\times}10^{-5}$ & 3.43706 & 674 \\
4 & GELU-MLP $[64]{\times}4$, $660\,$s & $20/20$ & 4.47999 & $1.3{\times}10^{-4}$ & 4.47992 & 676 \\
\bottomrule
\end{tabular}
\label{tab:stats}
\end{table}

\subsection{Failure modes in detail}
\label{app:modes}
Section~\ref{sec:free} sorts every unsuccessful uniform-protocol run into
one of three modes, and table~\ref{tab:failures} gives the totals. In this
appendix we state the rule that makes the assignment, give the counts for each
network, and describe the runs of each mode.

\paragraph{Classification rule.}
Let $B_{\rm t}$ be the target charge and $E_{\rm ref}$ the published
reference energy of table~\ref{tab:bounds}~\cite{battye2002,battye2001}. A run is assigned to mode~(a)
if $|B-B_{\rm t}|\ge10^{-3}$. A run with $|B-B_{\rm t}|<10^{-3}$ is
assigned to mode~(b) if $|E_2/E_4-1|\ge0.05$, that is, if it fails the
virial check of the verification protocol of section~\ref{sec:eval}.
Every remaining unsuccessful run passes the protocol and fails only the
$1\%$ energy requirement, $(E-E_{\rm ref})/E_{\rm ref}\ge0.01$, and is
assigned to mode~(c). No run with
$|B-B_{\rm t}|<10^{-3}$ fails the $\Delta E$ check or the bound
$E\ge|B|$ of the protocol.

\paragraph{Counts for each network.}
Table~\ref{tab:modes} gives
the resulting counts for each of the four networks. The success counts of table~\ref{tab:uniform}
are insensitive to the $1\%$ energy threshold: with $0.5\%$ in place of
$1\%$, no cell changes, and with $0.2\%$ the $B=3$ cell of
KAN$[16]{\times}4$ loses one run.

\begin{table}[h]
\centering
\caption{The failure modes of table~\ref{tab:failures}, listed for each
network. Each entry gives the number of runs in mode~(a), in mode~(b) and
in mode~(c), in that order. Each network contributes $10$ runs at each of
$B=1$ and $B=2$ and $20$ runs at each of $B=3$ and $B=4$.}
\begin{tabular}{lcccc}
\toprule
network & $B=1$ & $B=2$ & $B=3$ & $B=4$ \\
 & (a)/(b)/(c) & (a)/(b)/(c) & (a)/(b)/(c) & (a)/(b)/(c) \\
\midrule
KAN $[16]{\times}4$ & 0 / 0 / 0 & 0 / 0 / 0 & 1 / 0 / 0 & 6 / 1 / 2 \\
KAN $[21]{\times}4$ & 0 / 0 / 0 & 0 / 0 / 0 & 2 / 0 / 1 & 6 / 0 / 2 \\
GELU-MLP $[64]{\times}4$ & 0 / 0 / 0 & 0 / 0 / 0 & 1 / 0 / 3 & 5 / 1 / 4 \\
GELU-MLP $[48]{\times}4$ & 0 / 0 / 0 & 0 / 0 / 1 & 2 / 2 / 3 & 4 / 5 / 5 \\
\midrule
all four networks & 0 / 0 / 0 & 0 / 0 / 1 & 6 / 2 / 7 & 21 / 7 / 13 \\
\bottomrule
\end{tabular}
\label{tab:modes}
\end{table}

\paragraph{Mode (a): sector escape.}
Twenty-five of the twenty-seven runs in this mode end with an energy
between $1.5$ and $240$ times the reference energy, and with a virial ratio
$E_2/E_4$ that is off by more than $14\%$. The other two runs, both of
GELU-MLP$[64]{\times}4$, unwind all the way to the vacuum and end at
$B\approx0$ and $E\approx0$. Including these two runs, $11$ of the $27$ runs end with a $B$ whose
nearest integer differs from the target charge. To see how the charge
changes, we re-ran the two runs that unwind to the vacuum with the same settings and seeds,\footnote{These
two re-runs ran on an NVIDIA GeForce RTX 3060 GPU, not on the RTX 6000 Ada
GPU of section~\ref{sec:protocol}.}
and recorded the minimum of $\|\mathbf{v}\|$ on a grid of spacing $0.1$.
We recorded it at every epoch for the first $2\,000$ epochs, and at every
20th epoch after that. The re-runs agree with the records for the first few
epochs, then follow a different path and end in the target sector rather than
the vacuum. In
both re-runs the charge changed during training. When it
changed in the first $2\,000$ epochs, the minimum of $\|\mathbf{v}\|$ on
the grid usually fell to between $0.01$ and $0.09$, $4$ to $17$ times
below its median over the first $2\,000$ epochs. In the first $2\,000$ epochs, most changes of the charge therefore happened
when $\mathbf{v}$ came close to zero on the grid, as
section~\ref{sec:freezemech} describes.

\paragraph{Mode (b): unrelaxed configurations.}
In each of the nine runs in this mode the charge is within $10^{-3}$ of the
target charge, but the virial ratio is off by $8$ to $39\%$ and the energy
is $10$ to $39\%$ above the reference energy. In all nine runs the energy
was still falling when the $20\,000$-epoch budget ended: over the last $1\,000$
epochs it fell by between $0.025$ and $0.24$.
Section~\ref{sec:failmodes} reports what happens when the training is longer.

\paragraph{Mode (c): high energy.}
A mode-(c) run keeps the target charge and passes the protocol, and only
its energy shows that it is not the minimiser. To see how such a
configuration differs from the minimiser, we read its shape from the
principal moments of the inertia tensor~\eqref{eq:inertia}. The deviation
from the symmetry of the minimiser is measured as in
section~\ref{sec:free}: by $(I_2-I_1)/I_1$ at $B=2$ and by
$(I_3-I_1)/I_1$ at $B=3$ and $B=4$.
The single run at $B=2$ has
$(I_2-I_1)/I_1=0.29$ and lies $1.9\%$ above the reference energy.
At $B=3$ and $B=4$ the deviation is measured by the same quantity,
$(I_3-I_1)/I_1$, so the twenty runs at these charges are treated together.
The twenty runs have $(I_3-I_1)/I_1$ between $0.095$ and $4.1$, and lie
$1.2$ to $10.3\%$ above the reference energy. Over the twenty runs the
energy excess rises with $(I_3-I_1)/I_1$, with a Spearman rank correlation
of $+0.79$: the further a configuration is from the symmetry of the
minimiser, the higher its energy.

\subsection{Quadrature failure of the \texttt{FastKAN} variant}
\label{app:quadfail}
Some \texttt{FastKAN} runs pass the verification protocol but end with an
energy that is too low. This appendix shows that the quadrature
under-counts the energy of these runs.
\texttt{FastKAN} replaces the splines of the KAN layer by radial basis
functions (section~\ref{sec:arch}). No result of the main text uses a
\texttt{FastKAN} run. Of the $144$ \texttt{FastKAN} runs
(appendix~\ref{app:charge}), $72$ are at $B=1$ and $72$ at $B=2$. All runs
at $B=2$ end above the published reference energy of
table~\ref{tab:bounds}~\cite{battye2002,battye2001}. At $B=1$ the
minimiser is believed to be spherically symmetric, and the
one-dimensional hedgehog profile gives the energy $1.231442$ on $r\le60$
(section~\ref{sec:methods}). $16$ of the $72$ runs at $B=1$ end
below this value, by $2.4\times10^{-6}$ to $1.6\times10^{-4}$. However,
this shortfall does not mean that the fields of these runs have an energy
below $1.231442$. Although all $16$ satisfy the verification protocol of
section~\ref{sec:eval}, the following two checks show that the shortfall
comes from the quadrature. First, the charge is also below $1$ in all $16$
runs: $1-B$ reaches $1.6\times10^{-4}$, as large as the shortfall in the
energy. The energy and the charge come from the same quadrature, so both
are under-counted. The ratio $E/B$ corrects for the under-count, and lies
above $1.231442$ in all $16$.\footnote{This is the same correction that
Battye and Sutcliffe~\cite{battye2002,battye2001} apply to their raw grid
energies to obtain the published reference energies (point~(iii) of
section~\ref{sec:findings}).} Second, three of the $16$ configurations
were re-run and evaluated on a finer grid,
$80\times44\times60$. On this grid all three energies are
above $1.231442$. The likely cause of the shortfall is the following.
The quadrature sum evaluates both densities only at the nodes, and in some directions the field of \texttt{FastKAN} varies on a
scale finer than the spacing of the evaluation grid. The missing part of
both integrals is then likely to lie between the nodes.

At $B=1$ the anchored sweep consists of KAN and tanh-MLP runs
(section~\ref{sec:fair}). These runs show no
such under-count. One run of each network was re-run,
and its energies on the
$48\times28\times36$ and $80\times44\times60$ grids differ by less than
$3\times10^{-8}$. So, in these two runs, the $48\times28\times36$ grid
misses no energy between its nodes.

\section{Fairness of the comparison}
\label{app:fairapp}

\subsection{Fairness checklist}
\label{app:fair}
The outcome of a comparison between two architectures depends on the settings
that each architecture is given. Our conclusions could come from the settings that we chose, and
not from the KAN and the MLP themselves. To test this, we changed three
settings, one at a time, and left every other setting at the value used in
the main text. For
each test we give the purpose of the runs, their number and the result.

(i)~\emph{Learning rate.} Table~\ref{tab:fair} gives the energies of the
networks at the shared rate $2\times10^{-3}$ (row~1) and at the best of the
five rates for each network, which is $10^{-2}$ for all three (row~2). To
compare the networks at each rate, we added runs of the KAN$[16]{\times}4$ and both MLPs at
$[32]{\times}4$ at
$5\times10^{-4}$, $10^{-3}$, $5\times10^{-3}$ and $10^{-2}$ and at three
seeds each ($36$ runs). At each of these four rates the KAN has the lowest mean energy, and
the tanh-MLP has the highest.
KAN$[16]{\times}4$ has fewer parameters than the MLPs of
table~\ref{tab:fair}, $7\,920$ against $12\,996$. At each of the five
rates KAN$[16]{\times}4$ is lower than the tanh-MLP$[64]{\times}4$. At
each of these rates the mean energies of KAN$[16]{\times}4$ and the
GELU-MLP$[64]{\times}4$ differ by less than three standard errors of the
difference. The $36$ added runs are not used in table~\ref{tab:fair}.

(ii)~\emph{MLP activation.} Table~\ref{tab:fair} compares the tanh-MLP with
the GELU-MLP. To compare another smooth activation with tanh and GELU,
we also ran
SiLU at depth $4$ and widths $16$, $32$ and $64$, at three seeds each ($9$
runs). At each of the three widths, the mean energy of SiLU lies between
those of tanh and GELU. These $9$ runs are not used in
table~\ref{tab:fair}.

(iii)~\emph{Wall-clock budget, anchored track.} Rows $3$ and $4$ of
table~\ref{tab:fair} stop the training after $30\,$s and after $120\,$s
instead of after the $1000$ epochs of the anchored sweep. In both rows the
KAN is lower than the
tanh-MLP. To see whether the KAN is also lower at other sizes, we stopped
the KAN$[16]{\times}2$ and the tanh-MLP$[32]{\times}4$ after $30\,$s and
after $120\,$s, at three seeds each ($12$ runs). In these runs, the KAN is
again lower than the tanh-MLP after both times.
These two networks have about the same number of parameters, $3\,312$ and
$3\,428$.
We did not run the GELU-MLP at another size. These $12$ runs are not used
in table~\ref{tab:fair}.

\subsection{Implementation cost}
\label{app:implcost}
The KAN results of this paper all come from our own implementation
(section~\ref{sec:arch}). The outcome of a comparison between a KAN and an
MLP can change with how each network is implemented. To measure how the
implementation of the KAN affects the speed, we ran \texttt{pykan} (v0.2.8) with the networks $[4]{\times}3$ and
$[8]{\times}2$ at both $B=1$ and $B=2$, and with $[16]{\times}4$ at $B=2$, at
three seeds each ($15$ runs). We compared these runs with those of our
implementation for the same networks in the anchored sweep. The two sets
of runs use the same training settings, such as the number of epochs, the
training grid and the spline grid. Only the implementation differs. In this
comparison,
our implementation is $3$ to $24$ times faster. The difference comes from the code: \texttt{pykan} splits each
training step into many small operations, and the time that the CPU needs
to send the operations to the GPU limits the speed of \texttt{pykan}.

\section{Settings and run records}
\label{app:inventory}
Sections~\ref{sec:model} and~\ref{sec:methods} state the main settings of
both tracks. We give five more details here. (i)~The radial profile of
the anchored track is pre-optimised in the form
$f=(1-\xi)^p(\pi+\xi\,h(\xi))$ with $p=\tfrac12(1+\sqrt{1+8B})$ as in
section~\ref{sec:field} and $\xi=r/(2+r)$, where $h$ is a
tanh network of two layers of width $32$; the pre-optimisation uses
$1500$ Adam steps at learning rate $5\times10^{-3}$ and then $400$
iterations of the quasi-Newton method L-BFGS~\cite{liu1989}.
(ii)~The initial values of the spline coefficients of the KAN are drawn
from a normal distribution of mean zero and standard deviation
$0.1/\sqrt{n_{\rm in}}$, where $n_{\rm in}$ is the number of inputs
of the layer: $3$ for the first layer and the width $w$ for the others
(section~\ref{sec:arch}). (iii)~The weights of the MLP
use the PyTorch default. (iv)~The anchored track uses seeds $0$ to $4$,
and the check of row~2 of table~\ref{tab:fair} in section~\ref{sec:fair}
adds seeds $5$ to $9$ for the
KAN$[32]{\times}3$ at the learning rates $5\times10^{-3}$ and $10^{-2}$.
The free track uses seeds $0$ to $9$ at $B\le2$ and $0$ to $19$ at
$B\ge3$. (v)~In a run with a fixed number of epochs, the cosine schedule
lowers the learning rate to its minimum, $2\%$ of its initial value, at
the last epoch. However, in the runs that stop after a fixed time (rows~3 and~4 of
table~\ref{tab:fair}, table~\ref{tab:wallclock} and
appendix~\ref{app:fair}), the number of epochs is not known in advance,
so the cosine schedule is set during the run. The learning rate stays at
its initial value
for the first $10$ epochs. From the time of these $10$ epochs, we
extrapolate linearly to the epoch $N$ at which the time limit ($30\,$s,
$120\,$s, $160\,$s or $660\,$s) is reached. The cosine decay is set so
that the learning rate reaches this minimum at epoch $N$. If a run
reaches epoch $N$ before the time limit, the cosine continues, so the
learning rate rises again after epoch $N$.

Table~\ref{tab:inventory} summarises the $1\,398$ training runs behind the
numbers reported in this paper.
\begin{table}[h]
\centering
\caption{Training runs that this paper reports, grouped by the experiment
that produced them. Every run is stored with its settings, seed, history
and verification metrics. In each row, the parentheses list the places
where the paper reports that experiment. The rows are grouped by the first of
these places, which is named in the heading above them.}
\begin{tabular}{lr}
\toprule
Experiment & runs \\
\midrule
\textit{Section~\ref{sec:fair}} & \\
Anchored sweep (section~\ref{sec:fair}, figure~\ref{fig:pareto}, table~\ref{tab:fair}, appendices~\ref{app:charge} and~\ref{app:quadfail}) & 600 \\
Table~\ref{tab:fair}, row~2: learning rates $5\times10^{-4}$, $10^{-3}$, $5\times10^{-3}$ and $10^{-2}$, three networks, five seeds (section~\ref{sec:fair}) & 60 \\
Table~\ref{tab:fair}, rows~3 and~4: stopped after $30\,$s and after $120\,$s, three networks, five seeds (section~\ref{sec:fair}) & 30 \\
Seeds $5$ to $9$ of KAN$[32]{\times}3$ at learning rates $5\times10^{-3}$ and $10^{-2}$ (section~\ref{sec:fair}; appendix~\ref{app:inventory}~(iv)) & 10 \\
Re-runs of the $60$ runs behind table~\ref{tab:fair}, evaluated also on the $80\times44\times60$ grid (section~\ref{sec:fair}) & 60 \\
\midrule
\textit{Section~\ref{sec:free}} & \\
Uniform protocol: KAN$[16]{\times}4$, KAN$[21]{\times}4$, GELU-MLP$[64]{\times}4$ and GELU-MLP$[48]{\times}4$ (table~\ref{tab:uniform}) & 240 \\
Equal wall-clock budget: KAN$[16]{\times}4$ for $160\,$s and GELU-MLP$[64]{\times}4$ for $660\,$s (section~\ref{sec:wallclock}, table~\ref{tab:wallclock}) & 80 \\
\midrule
\textit{Section~\ref{sec:freeze}} & \\
The sixteen runs of table~\ref{tab:freezeruns} (subsection~\ref{sec:freezeres}) & 16 \\
The imprinted run of figure~\ref{fig:freeze} (subsection~\ref{sec:freezeres}) & 1 \\
Runs that vary how far the imprinting regression is taken, at $B_{\rm t}=2$ (subsection~\ref{sec:imprintres}, table~\ref{tab:imprintdesign}) & 50 \\
\midrule
\textit{Section~\ref{sec:discussion}} & \\
KAN$[16]{\times}4$ and KAN$[8]{\times}2$ at $G=3$ and $G=10$, three seeds (section~\ref{sec:limitations}) & 12 \\
Thirteen re-runs to measure GPU nondeterminism (section~\ref{sec:repro}) & 13 \\
\midrule
\textit{Appendix~\ref{app:verif}} & \\
Anchored-track \texttt{FastKAN} runs (appendices~\ref{app:charge} and~\ref{app:quadfail}) & 144 \\
Re-runs of the two runs that unwind to the vacuum, recording the minimum of $\|\mathbf{v}\|$ (appendix~\ref{app:modes}) & 2 \\
Re-runs of five anchored runs at $B=1$, evaluated also on the $80\times44\times60$ grid (appendix~\ref{app:quadfail}) & 5 \\
\midrule
\textit{Appendix~\ref{app:fairapp}} & \\
KAN$[16]{\times}4$, tanh-MLP$[32]{\times}4$, GELU-MLP$[32]{\times}4$: four learning rates, three seeds (appendix~\ref{app:fair}~(i)) & 36 \\
MLP with SiLU activation at depth $4$ and widths $16$, $32$ and $64$, three seeds (appendix~\ref{app:fair}~(ii)) & 9 \\
KAN$[16]{\times}2$ and tanh-MLP$[32]{\times}4$ stopped after $30\,$s and after $120\,$s, three seeds (appendix~\ref{app:fair}~(iii)) & 12 \\
Anchored-track \texttt{pykan} runs (appendix~\ref{app:implcost}) & 15 \\
\midrule
\textit{Appendix~\ref{app:relax}} & \\
Re-runs of three runs of table~\ref{tab:uniform}, saving the baryon density during training (appendix~\ref{app:relax}) & 3 \\
\midrule
Total & 1\,398 \\
\bottomrule
\end{tabular}
\label{tab:inventory}
\end{table}

\section{Formation of the symmetric shapes during relaxation}
\label{app:relax}
\begin{figure}[h]
\centering
\includegraphics[width=\textwidth]{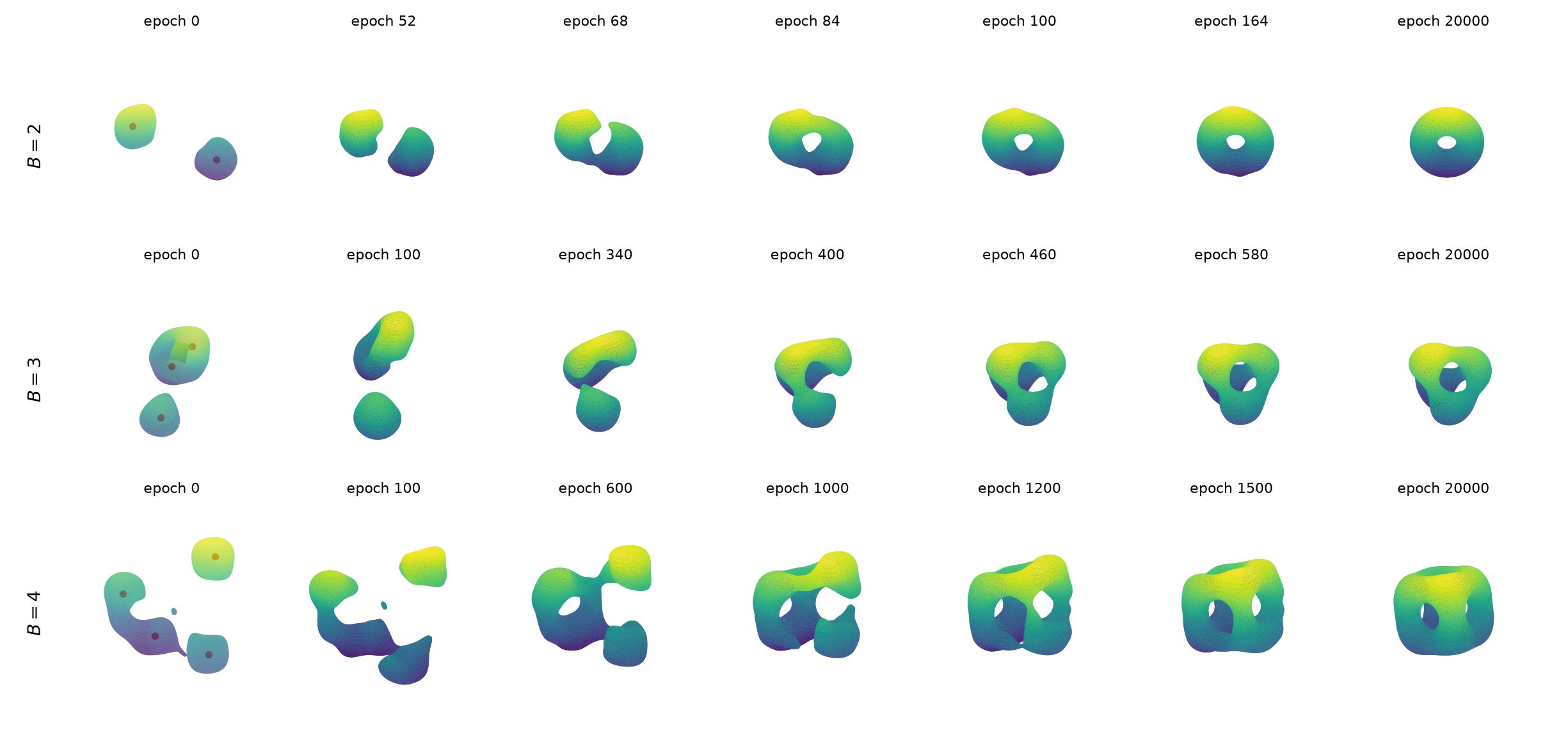}
\caption{Baryon-density isosurfaces during training at $B=2$, $3$ and $4$ (rows). The epoch is given above each panel. All panels of a row use the same isosurface level: $50\%$ of the maximum baryon density in the last panel (epoch $20\,000$) of that row. At epoch $0$ the dots mark the centres of the imprinted lumps (section~\ref{sec:imprint}); at $B=3$ and at $B=4$, two of the lumps form one surface. The surfaces at epoch $0$ are drawn translucent so that the dots are visible.}
\label{fig:relax}
\end{figure}

Figure~\ref{fig:relax} shows the baryon-density isosurface at several epochs, from the imprinted lumps at epoch $0$ to the shapes of figure~\ref{fig:iso}. The runs start from the same imprinted lumps as the headline solutions; at $B=3$ and $B=4$ they are trained for $20\,000$ epochs instead of the about $80\,000$ epochs of the headline solutions. We re-ran the runs to save the field during training, and the re-runs reproduce the recorded energy and charge exactly.

\end{document}